\documentclass[%
a4paper,
superscriptaddress,
notitlepage,
bibnotes,
 amsmath,amssymb,
 aps,
10pt,
twocolumn,
pre,
]{revtex4-1}

\usepackage{mathtools}

\usepackage{xcolor}
\colorlet{mylinkcolor}{blue!66!black!80}

\usepackage[colorlinks=true,linkcolor=mylinkcolor,citecolor=mylinkcolor,filecolor=cyan,urlcolor=mylinkcolor,breaklinks=true]{hyperref}

\usepackage[utf8]{inputenc}

\newcommand{\mat}[1]{\textbf{#1}}
\newcommand{\avg}[1]{\langle#1\rangle}
\newcommand{\bavg}[1]{\big\langle#1\big\rangle}
\newcommand{\bbavg}[1]{\Big\langle#1\Big\rangle}
\newcommand{\del}{\partial}
\newcommand{\dd}{{\rm d}}

\newcommand{\bbb}{\boldsymbol{b}}

\newcommand{\bs}{\boldsymbol{s}}

\newcommand{\by}{\boldsymbol{y}}
\newcommand{\bz}{\boldsymbol{z}}
\newcommand{\bmu}{\boldsymbol{\mu}}

\newcommand{\x}{{\rm x}}
\newcommand{\y}{{\rm y}}
\newcommand{\e}{{\rm e}}

\newcommand{\m}{\mathrm{m}}
\newcommand{\bSigma}{\boldsymbol{\Sigma}}

\newcommand{\T}{\top}

\newcommand{\eq}{\mathrm{eq}}

\begin{document}
\title{Optimal inference strategies and their implications for the linear noise approximation}
\author{David Hartich}
\affiliation{%
II. Institut für Theoretische Physik, Universität Stuttgart, 70550 Stuttgart, Germany}
 \author{Udo Seifert}%
\affiliation{%
II. Institut für Theoretische Physik, Universität Stuttgart, 70550 Stuttgart, Germany}

\begin{abstract}
We study the information loss of a class of inference strategies that is solely based on time averaging.
For an array of independent binary sensors (e.g., receptors, single electron transistors) measuring a weak random signal (e.g., ligand concentration, gate voltage) this information loss is up to 0.5\,bit per measurement irrespective of the number of sensors.
We derive a condition related to
the local detailed balance relation that determines whether or not such a loss of information occurs. Specifically, if the free energy difference arising from the signal is symmetrically distributed among the forward and backward rates, time integration mechanisms
will capture the full information about the signal. As an implication, for the linear noise approximation,
we can identify the same loss of information, arising from its inherent simplification of the dynamics.

\end{abstract}

\maketitle

\section{Introduction}
In their seminal work \cite{berg77}, Berg and Purcell (BP) have explored the fundamental limits to the accuracy with which
cellular organisms can perform concentration measurements with receptors via time averaging.
Later, it has then been discovered that a receptor performing a maximum likelihood (ML) measurement can reduce the uncertainty of a measurement even further by a factor of two \cite{endr09}. In an information theoretic language the ML estimate ``uses'' 0.5\,bit more information from the history of the receptor occupancy than the BP estimate.
Further studies have examined ML estimates for concentration ramps \cite{mora10,aqui14}, for spatial gradients \cite{hu10}, and
for competing signals \cite{mora15} (see \cite{aqui16,wold16} for recent reviews). Most prominently, it has been elucidated that this additional information cannot be reached within the class of linear models \cite{gove12,meht12},
which illustrates the inherent loss of information for this specific class of inference strategies.

Thermodynamic constraints for the information acquisition of biological networks
have attracted much interest recently
\cite{tu08a,lan12,meht12,bara13a,palo13,skog13,lang14,gove14,gove14a,bara14b,sart14,hart15,bo15,ito15,bara15,bara15a,wold16,meht16,lan16}. Specifically, chemical networks may only reach the highest sensory performance at diverging thermodynamic costs \cite{meht12,lang14,gove14,gove14a}. Using the optimal strategy to be able to come close to such
bounds is, therefore, of fundamental importance.

In this paper, we will show how the maximum sensory performance depends on the strategy
used for inferring the signal. Thereby, we compare different classes of inference strategy, for example time averaging and
counting transitions \cite{bara15a} as well as their joint correlations. For binary sensors
responding to a signal, we illustrate how the thermodynamic local detailed balance relation
determines the sensory limits for time averaging mechanisms but does not generally determine the maximal information attainable by counting transitions.
In particular, we will show under which conditions time averaging mechanisms loose information (e.g., up to 1/2 bit per measurement).
The same loss of information can also be identified in the linear noise approximation, which is a powerful tool to approximate large chemical networks with an effective Brownian motion \cite{kamp07,elf03,tkac11,bres14,horo15a}. We will show under which circumstances these approximation schemes
are almost accurate on the trajectory level and when they loose information up to about $1/2\,$bit.

The paper is organized as follows. In Sec. \ref{sec:model}, we define the model system of a single binary sensor measuring a stochastic signal. In Sec. \ref{sec:main_results}, we derive
a thermodynamic expression Eq. \eqref{eq:def_dI} for the information loss of inference strategies
that are solely based on time averaging. In Sec. \ref{sec:LNA}, we discuss the implication for approximation schemes that involve a continuous Brownian motion (linear noise approximation for weak signals). We conclude in Sec. \ref{sec:conclusion}.
Appendix \ref{sec:information_vs_error} shows basic relations between information and error.
In Appendix \ref{sec:gen_func} we introduce the generating function that can be used to calculate all quantities of interest.
Appendix \ref{sec:robust} explains the robustness of our main result, Eq. \eqref{eq:def_dI}, against non-linearities arising from strong signals.
A comparison to  Refs. \cite{bara15a,koza99,wach15} can be found in Appendix \ref{sec:method}.
In Appendix \ref{sec:filter}, we calculate the mutual information for coarse-grained processes
within the linear noise approximation.

\begin{figure}%
\includegraphics{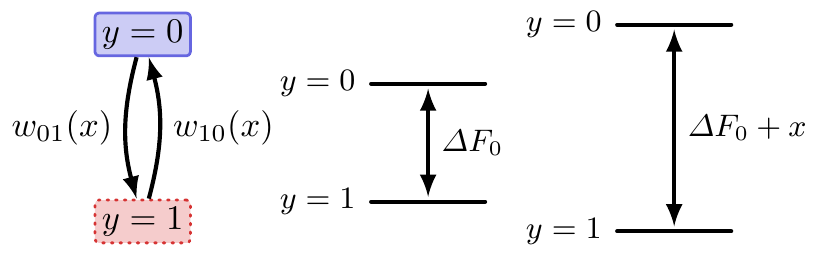}
\caption{The two level system with transition rates is shown on the left.
The change $x$ of the free energy difference between the the states due to the signal is shown
on the right.}%
\label{fig:level}%
\end{figure}

\section{Model}
\label{sec:model}
We assume a sensor $y$ which, at time $t$, can be in two different states, the ``empty state'' $y_t=0$ or the ``occupied state'' $y_t=1$. These states are initially separated by a free energy difference $\varDelta F_0$. The signal then changes the free energy difference to $\varDelta F_0+x$ such that a positive signal ($x>0$) favors the occupied state $y=1$. More precisely, as illustrated in Fig. \ref{fig:level},
the transition rates $w_{yy'}(x)$ from $y_t=y$ to $y_t=y'$ satisfy the local detailed balance relation \cite{seif12}
\begin{equation}
\ln\frac{w_{01}(x)}{w_{10}(x)}= \varDelta F_0 +x,
 \label{eq:LDB}
\end{equation}
where we set $k_{\rm B}T\equiv 1$ throughout.
In equilibrium, the probability of being in state $y=1$ is then given by $P^\eq(y{=}1|x)=\e^{\varDelta F_0 +x}/[\e^{\varDelta F_0 +x}+1]$.
For an illustration, let us consider the paradigmatic case of a single receptor measuring a concentration $c$, where the binding of a ligand to the receptor
occurs with a rate $w_{01}\equiv kc$ and ligands are released from the receptor with a rate $w_{10}=r$. One can then identify $\varDelta F_0=\ln (kc_0/r)$ and $x=\ln(c/c_0)$, where $c_0$
is some reference concentration,
i.e., $x$ corresponds to a change of the external chemical potential. The mean occupation level then becomes $P^\eq(1|x)=kc/(kc+r)$. We will use this example of a receptor measuring a concentration later, since it allows us to compare our results with the findings in, e.g., Refs. \cite{endr09,mora10,gove12,meht12}. As a non-biological example, one can consider as sensor
a single electron transistor \cite{kosk14a,kosk14,kosk15}, trying to infer a signal $x$, which is a change of gate voltage \cite{kosk14}.

The local detailed balance relation \eqref{eq:LDB} provides one constraint for two rates.
Hence, it does not generally determine both rates $w_{01}(x)$ and $w_{10}(x)$ individually.
The asymmetry parameter $\alpha$, a central quantity of this paper, defined through
\begin{equation}
\begin{aligned}
  \alpha&\equiv -\del_x\ln w_{10}(x)=1-\del_x\ln w_{01}(x)
\end{aligned}
\label{eq:def_alpha}
\end{equation}
accounts for this freedom of choice. For ``normal'' values $0\le\alpha\le1$, the signal $x$ influences the rates in such a way that one rate increases while the respective reverse rate decreases.
For $\alpha=1/2$ the signal $x$ has a symmetric influence on both rates, whereas for the above example with the receptor we have $\alpha=0$, i.e., only one rate is affected by the signal,
since, $w_{01}(x)=k c_0\e^x$ and $w_{10}(x)=r$.
Moreover, for fermionic rates as relevant to the single electron box $w_{01}(x)\equiv \gamma P^{\eq}(1|x)$ and $w_{01}(x)\equiv\gamma[1- P^{\eq}(1|x)]$ (see, e.g., Ref. \cite{stra13}), the asymmetry parameter becomes $\alpha=P^{\eq}(1|x)$.
Similarly, in the experiment reported in \cite{kosk14a}, $\alpha\simeq 1/2$
corresponds to a single electron box for which the mean number of electrons is
about 1/2, as can be deduced from the supporting information (Fig. S1) of Ref. \cite{kosk14a}.

\section{Main results}
\label{sec:main_results}
We first define the measurement problem as illustrated in Fig. \ref{fig:level}. We assume the sensor is initially equilibrated with free energy difference $\varDelta F_0$ at time $t<0$.
At $t=0$ the signal $x$ changes the free energy difference to $\varDelta F_0+x$, where we assume that $x$ is normally distributed
with zero mean and variance $\mathcal{E}_\x^2$, i.e., $P(x)=(2\pi \mathcal{E}_\x^2)^{-1/2}\exp[-x^2/(2\mathcal{E}_\x^2)]$. For weak signals the condition $\mathcal{E}_{\mathrm{x}}^2\ll1$ holds.
The goal is to determine the mutual information that the time series $\{y_t'\}_{0\le t'\le t}$ of the sensor
contains about the value $x$. Since there is a direct relation between estimation error and information \cite{cove06}, see also Appendix \ref{sec:information_vs_error}, the former can be inferred from the latter.
Moreover, there is a direct link between mutual information and thermodynamics \cite{parr15}.
The mutual information is given by \cite{cove06}
\begin{equation}
 \mathcal{I}\equiv I[x:\{y_{t'}\}_{0\le t'\le t}] \equiv\bbavg{\ln\frac{P(\{y_{t'}\}_{0\le t'\le t}|x)}{P(\{y_{t'}\}_{0\le t'\le t})}},
 \label{eq:def_I}
\end{equation}
where $P(\{y_{t'}\}_{0\le t'\le t}|x)$ is the conditional probability of the sensor trajectory given the signal value and $P(\{y_{t'}\}_{0\le t'\le t})$
is its marginal distribution. The brackets $\avg{{\cdots}}$ denote the average over all possible
realizations $\{y_{t'}\}_{0\le t'\le t}$ and $x$, weighted by the corresponding joint distribution $P(\{y_{t'}\}_{0\le t'\le t}|x)P(x)$. We will later use $\avg{{\cdots}}_x$ for the conditional average over $\{y_{t'}\}_{0\le t'\le t}$ weighted by $P(\{y_{t'}\}_{0\le t'\le t}|x)$.
The stochastic time evolution of a discrete Markov process in continuous time consists of a sequence of exponential decays. Hence, the
path probability reads (see, e.g., \cite{endr09,seif12})
\begin{multline}
 P(\{y_{t'}\}_{0\le t'\le t}|x)\equiv \\
 P(y_0)w_{01}(x)^{\hat{n}_{01}}w_{10}(x)^{\hat{n}_{10}}\exp[-w_{01}(x)\hat{\tau}_0-w_{10}(x)\hat{\tau}_1]
 \label{eq:def_pathprob}
\end{multline}
where $\hat{n}_{yy'}$ is the number of transitions from $y$ to $y'$ and $\hat{\tau}_y$ is the time spent in state $y$ along the trajectory $\{y_{t'}\}_{0\le t'\le t}$.
Since $\hat{\tau}_0=t-\hat{\tau}_1$ and, since each jump $y=0\to y=1$ must be followed by the reverse transition, the path weight \eqref{eq:def_pathprob} is fully determined through $y_0, \hat{n}_\mathrm{tot}\equiv \hat{n}_{01}+\hat{n}_{10}$ and $\hat{\tau}_1$, i.e.,
 $\mathcal{I}=I[x{:}y_0,\hat{\tau}_1,\hat{n}_\mathrm{tot}]$.

We now discuss the important role of the asymmetry parameter $\alpha$ from \eqref{eq:def_alpha} for this information acquisition problem in the
long time limit. For $t\to\infty$, the total number of jumps satisfies $\hat{n}_\mathrm{tot}/2\simeq \hat{n}_{01}\simeq \hat{n}_{10}\gg1$.
Most prominently, for $\alpha=1/2$ the first terms of the path probability \eqref{eq:def_pathprob}
do not depend on the signal, i.e., $ w_{01}(x)^{\hat{n}_{01}}w_{10}(x)^{\hat{n}_{10}}\simeq
[w_{01}(0)w_{10}(0)]^{\hat{n}_\mathrm{tot}/2}$. In this specific case, the total number of jumps $\hat{n}_\mathrm{tot}$ does not contain additional information about $x$. Consequently,
the coarse grained statistics $(y_0,\hat{\tau}_1)$ that contains the information
\begin{equation}
\tilde{\mathcal{I}}\equiv
I[x{:}y_0,\hat{\tau}_1]=\bbavg{\ln\frac{P(y_0,\hat{\tau}_1|x)}{P(y_0,\hat{\tau}_1)}}
\label{eq:Itilde}
 \end{equation}
 about the signal, leads to $\tilde{\mathcal{I}}=\mathcal{I}$ for $\alpha=1/2$, whereas, in general, it satisfies the inequality $\tilde{\mathcal{I}}\le\mathcal{I}$.
\begin{figure}%
 \centering
\includegraphics{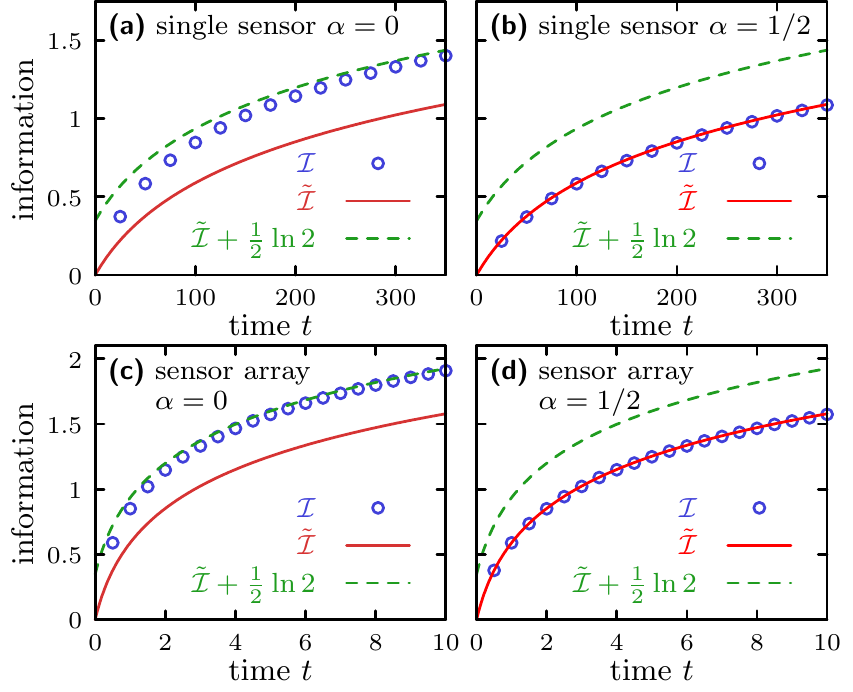}%
 \caption{Full information of trajectory (blue open circle) versus coarse grained information without knowledge of the number of jumps (red solid line).
 The results are shown for four different sensory systems, for which
the rates are either asymmetrically $\alpha=0$ (a),(c) or the symmetrically
$\alpha=1/2$ (b), (d) influenced  by the signal $x$.
The upper panel (a),(b) shows the single sensor case ($N=1$). The lower panel (c),(d) displays the results
for a sensor array with $N=100$ binary sensors, where $\mathcal{I}=\mathcal{I}_N$ and $\tilde{\mathcal{I}}=\tilde{\mathcal{I}}_N$.
$\mathcal{I}$ is obtained numerically by simulating $10^{6}$ individual trajectories
to estimate \eqref{eq:def_I} [or \eqref{eq:def_I_N}]. For the coarse grained mutual information $\tilde{\mathcal{I}}$ we use the approximation from
Eq. \eqref{eq:Itilde_N}. Parameters: Rates $w_{01}(x)\equiv\smash{\e^{(1-\alpha)x}}$, $w_{10}(x)\equiv\smash{\e^{-\alpha x}}$, $\alpha=0$ (left panel), $\alpha=1/2$ (right panel), $N=1$ (upper panel), $N=100$ (lower panel), signal standard deviation $\mathcal{E}_\x\equiv0.3$ ($\mathcal{E}_\x^2=0.09\ll1$), $p_0=1/2$, and $\omega_\y=2$.}%
\label{fig:4plots}%
\end{figure}%
In Fig. \ref{fig:4plots}(a),(b), we compare the full information $\mathcal{I}$ from the trajectory (blue open circles) with the information $\tilde{\mathcal{I}}$ that excludes the knowledge from the number of jump events (red solid line).
For symmetric weight ($\alpha=1/2$), we find $\mathcal{I}=\tilde{\mathcal{I}}$, i.e., the number of jump events
$\hat{n}_\mathrm{tot}$ does not provide any additional information about the signal.
Remarkably, for the asymmetric weight ($\alpha\neq 1/2$), as it applies to the receptor model discussed above ($\alpha=0$), the number of jump events $\hat{n}_\mathrm{tot}$ contributes to an additional amount of information up to $\frac{1}{2}\ln 2=1/2\,\mathrm{bit}$. The same result is obtained for an array of $N=100$ receptors [see Fig. \ref{fig:4plots}(c),(d)].
Since the measurement error (variance) is proportional to $\e^{-2\mathcal{I}}$ (see, e.g., \cite{cove06} or Appendix \ref{sec:information_vs_error}) this additional information
corresponds to an error reduction by a factor of 2.
This enhanced measurement accuracy for ML estimates was first found in \cite{endr09} for the above discussed specific receptor model ($\alpha=0$). We conclude that for binary sensors measuring
a signal an improved accuracy occurs whenever the signal has a non-symmetric
impact on the transition rates, i.e., for $\alpha\neq1/2$.
Note that each subfigure in Fig. \ref{fig:4plots} shows results for distinct models
or sensory systems.

Using the method of generating function, see Appendix \ref{sec:gen_func}, we obtain more generally
\begin{equation}
\varDelta\mathcal{I}\equiv\lim_{t\to\infty}(\mathcal{I}-\tilde{\mathcal{I}})=\frac{1}{2}\ln(4\alpha^2-4\alpha+2).
\label{eq:def_dI}
\end{equation}%
We present a simple derivation of this formula for weak signals $\mathcal{E}_\x^2\ll 1$ in
Appendix \ref{sec:gen_func}. In Appendix \ref{sec:robust}
we show that \eqref{eq:def_dI} holds even for strong and/or non-Gaussian signals.
This relation constitutes our first main result, namely, the asymmetry parameter
$\alpha$ significantly determines the additional information content of the number of jump events. For $0\le\alpha\le1$, the number of jump events
contributes to an additional information up to 1/2\,bit,
which can qualitatively be understood as follows. For a symmetric effect of the signal $x$
on the rates ($\alpha=1/2$), one rate increases and its reverse rate decreases
such that the overall number of transitions $\hat{n}_\mathrm{tot}$ remains the same
without establishing correlations with $x$, whereas in the extreme case, e.g., $\alpha=0$, only the transition $w_{01}(x)$ is affected by the signal resulting in a monotonic correlation between $\hat{n}_\mathrm{tot}$ and $x$.

From a more technical point of view, we note that the method of generating functions necessary to derive our main result, Eq. \eqref{eq:def_dI}, can also be used to describe in finite time the
joint dispersion of different classes of random variables, as for example, jump variables $\hat{n}_{yy'}$ and  persistence time $\hat{\tau}_y$ \cite{bara15a}.
We explain in Appendix \ref{sec:method} how one can adopt the methods from \cite{koza99,bara15a,wach15} to such a joint dispersion
in the long time limit.

The expression for the information gain $\varDelta\mathcal{I}$ \eqref{eq:def_dI} holds even for an arbitrary number of sensors,
which we show by generalizing the path weight \eqref{eq:def_pathprob}.
We denote for the $i$-th sensor the time spent in state $y$ by $\tau_y^{(i)}$ and the number of transitions
from  $y$ to $y'$ by $\hat{n}_{yy'}^{(i)}$, which results in the conditional path weight
\begin{multline}
 P(\{\by_{t'}\}_{0\le t'\le t}|x)\equiv p_0^{Y_0}(1-p_0)^{N-Y_0} \\
\times w_{01}(x)^{\hat{n}_{01}}w_{10}(x)^{\hat{n}_{10}}\exp[-w_{01}(x)\hat{\tau}_0-w_{10}(x)\hat{\tau}_1]
 \label{eq:def_N_pathprob}
\end{multline}
where $p_0\equiv \e^{\varDelta F_0}/(1+\e^{\varDelta F_0})$, $Y_0\equiv \sum_{i=1}^Ny_{t}^{(i)}|_{t=0}$,
$\tau_1\equiv\sum_{i=1}^N\tau_1^{(i)}$, $\tau_0= Nt-\tau_1$ and $\hat{n}_{yy'}\equiv\sum_{i=1}^N\hat{n}_{yy'}^{(i)}$.
The marginal path weight is given by $ P(\{\by_{t'}\}_{0\le t'\le t})= \int\dd xP(\{\by_{t'}\}_{0\le t'\le t}|x)P(x)$.
The full information between the time series of an array of $N$
sensors and the random signal $x$
then formally reads
\begin{equation}
 \mathcal{I}_N\equiv\bbavg{\ln\frac{P(\{\by_{t'}\}_{0\le t'\le t}|x)}{P(\{\by_{t'}\}_{0\le t'\le t})}},
  \label{eq:def_I_N}
\end{equation}
which can also be written in the form $\mathcal{I}_N=I[x{:} Y_0,\hat{\tau}_1,\hat{n}_{01},\hat{n}_{10}]$ using \eqref{eq:def_N_pathprob}. Note that $0\le\tau_1\le Nt$ holds in general.
The mutual information between signal and the coarse-grained statistics $(Y_0,\hat{\tau}_1)$
then formally generalizes to
\begin{equation}
 \tilde{\mathcal{I}}_N\equiv I[x{:}Y_0,\hat{\tau}_1].
 \label{eq:def_Itilde_N}
\end{equation}
Remarkably, in the long time limit $t\to\infty$, we find the same result as for the single
sensor case \eqref{eq:def_dI}
\begin{equation}
 \varDelta\mathcal{I}_N\equiv\lim_{t\to\infty}(\mathcal{I}_N-\tilde{\mathcal{I}}_N)=\frac{1}{2}\ln(4\alpha^2-4\alpha+2),
 \label{eq:dI_N}
\end{equation}
which holds for arbitrarily distributed signals as shown in Appendices
\ref{sec:gen_func} and \ref{sec:robust}.
This result agrees with Fig. \ref{fig:4plots}, where we observe the same
loss of information between $\alpha=0$ (asymmetric weight) and $\alpha=1/2$ (symmetric weight)
for a single sensor as well as for an array of sensors with $N=100$.

Assuming a weak signal as it applies to the results shown in Fig. \ref{fig:4plots}, we can obtain
\eqref{eq:dI_N} in a simpler way by using the following approximations. 
As shown in Appendix \ref{sec:gen_func} for weak signals ($\mathcal{E}_\x^2\ll1$)
the approximations
\begin{equation}
 \mathcal{I}_N\approx\frac{1}{2}\ln\left[1+(2\alpha^2-2\alpha+1)p_0(1-p_0)\mathcal{E}_\x^2N\omega_\y t \right],
 \label{eq:I_N}
\end{equation}
and
\begin{equation}
 \tilde{\mathcal{I}}_N\approx\frac{1}{2}\ln\left[1+\frac{1}{2}p_0(1-p_0)\mathcal{E}_\x^2N\omega_\y t \right],
 \label{eq:Itilde_N}
\end{equation}
hold, where $\omega_\y\equiv w_{01}(0)+w_{10}(0)$. The approximations are fairly accurate for $\mathcal{E}_\x=0.3$ as can be deduced from Fig. \ref{fig:4plots}(b),(d); see also Fig. \ref{fig:0.3vs2.0}.(a) in Appendix \ref{sec:robust}.
Comparing the approximations \eqref{eq:I_N} and \eqref{eq:Itilde_N} immediately leads to
\eqref{eq:dI_N}.
It should be  that the additional information $\varDelta \mathcal{I}$ can be also be reached
with a longer observation time $t\to t\exp(2\varDelta\mathcal{I})$, since $\tilde{\mathcal{I}}_N\approx\text{const.}+\frac{1}{2}\ln (Nt)$ for $t \to\infty$.

\section{Continuous approximation schemes}
\label{sec:LNA}
\begin{figure}%
 \centering%
\includegraphics{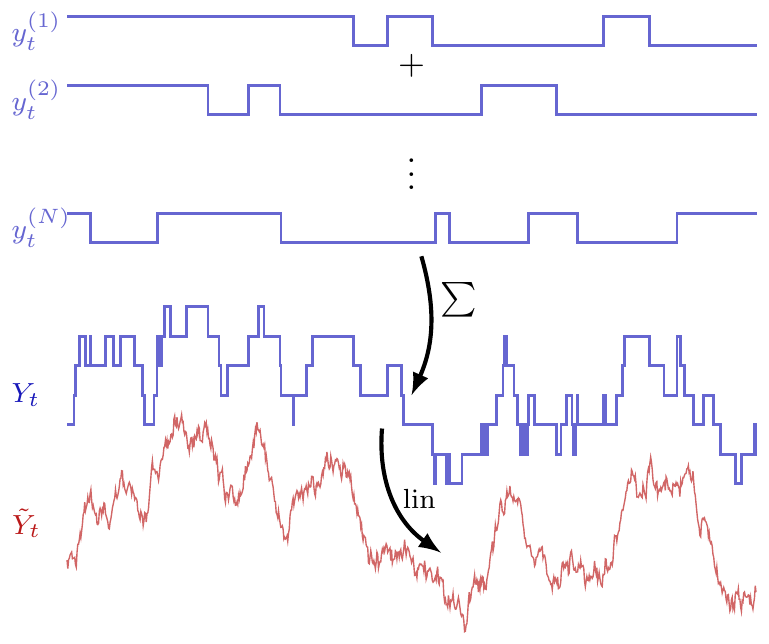}
 \caption{Two levels of coarse graining. First, $\by_t\equiv(y^{(1)}_t,y^{(2)}_t,\ldots,y^{(N)}_t)\to Y_t=Y[\by(t)]$. Second, a linear noise approximation
 $Y_t\to\tilde{Y}_t$.}%
\label{fig:sampletraj}%
\end{figure}

Complex chemical networks, for example the signaling networks in individual cells, typically contain of the order of $N=10^4\gg 1$ signaling molecules or receptors.
For such systems, it is often reasonable not to try to resolve each reaction of any individual component, but rather to use an approximation with deterministic rate equations or with a stochastic Brownian motion, see e.g., \cite{kamp07,elf03,tkac11,bres14,horo15a,pole15}.
For the present inference problem, the approximation of the system with a stochastic Brownian motion \cite{kamp07,elf03,tkac11,bres14,horo15a}
is especially interesting, since it considers fluctuations arising from the discreteness of the system.
We will see that this approximation scheme, as illustrated in Fig. \ref{fig:sampletraj},
looses information about the signal $x$ compared to the original system that remarkably is precisely given by \eqref{eq:def_dI} as well.

Before applying the approximation one
has to consider the total sensor activity $Y_t\equiv \sum_{i=1}^Ny^{(i)}_t$ of the sensor array.
Since the sensors are operating independently, one can verify that $\mathcal{I}_N=I[x{:}\{Y_{t'}\}_{0\le t'\le t}]$,
i.e., $\{Y_{t'}\}_{0\le t'\le t}$ contains the same amount of information about the signal $x$ as $\{\by_{t'}\}_{0\le t'\le t}$.
The linear noise approximation $\tilde{Y}_t$ is now used to approximate $Y_t\simeq \tilde{Y}_t$ with a Langevin equation. The approximation, sketched in Fig. \ref{fig:sampletraj}, is typically quite accurate for $N\gtrsim 10$ and weakly changing signals $x$ ($\mathcal{E}_\x^2\ll 1$).
One can identify three core properties that determine the dynamics of $\tilde{Y}_t$.
First, the correct mean behavior requires $\avg{\tilde{Y}_\infty}_x\approx \avg{Y_\infty}_x= Nw_{01}(x)/[w_{01}(x)+w_{10}(x)]$. A series expansion of the right hand side up to first order yields
\begin{equation}
 \avg{\tilde{Y}_\infty}_x=Np_0+Np_0(1-p_0)x
\end{equation}
where we have used $p_0= w_{01}(0)/[w_{01}(0)+w_{10}(0)]$ and \eqref{eq:LDB}. Second, $\avg{Y_{t}}_x$ decays exponentially
to $\avg{Y_\infty}_x$ with rate $\omega_\y\equiv w_{01}(0)+w_{10}(0)$ that also characterizes the relaxation speed of the system. Third, the steady state variance satisfies $\avg{Y_\infty^2}_x-\avg{Y_\infty}_x^2\approx Np_0(1-p_0)$.
Adopting these three properties leads to the effective Langevin equation \cite{bres14}
\begin{equation}
\dot {\tilde{Y}}_t=-\omega_\y\left[\tilde{Y}_t-\avg{\tilde{Y}_\infty}_x\right]+\xi_t,
\label{eq:langevin1}
\end{equation}
where  $\xi_t$ is standard white noise with  $\avg{\xi_t}_x=0$ and  $\avg{\xi_t\xi_{t'}}_x=2D\delta(t-t')$ with noise strength $D\equiv N\omega_\y p_0(1-p_0)$.
    
The process \eqref{eq:langevin1} is the well known linear Ornstein-Uhlenbeck process.
For such a linear process optimal filtering strategies exist, known as Kalman-Bucy filters, that infer a normally distributed stochastic value $x\sim \mathcal{N}(0,\mathcal{E}_\x^2)$ from the observation $\{\tilde{Y}_{t'}\}_{0\le t'\le t}$ with least variance $\tilde{\mathcal{E}}_t^2=2\mathcal{E}_\mathrm{x}^2/[2+\mathcal{E}_\mathrm{x}^2D t]$; see e.g. \cite{okse03,horo14a,hart16} or Appendix \ref{sec:filter}. Hence, due to
the Gaussian nature of the process this linearly approximated process contains information about the signal, which is given by
\begin{align}
\mathcal{I}^\mathrm{lin}_N&\equiv I(x{:}\{\tilde{Y}_{t'}\}_{0\le t'\le t})=\frac{1}{2}\ln\frac{\mathcal{E}_\x^2}{\tilde{\mathcal{E}}_t^2}
\nonumber\\
&=\frac{1}{2} \ln\left[1+\frac{1}{2}\mathcal{E}_\x^2p_0(1-p_0)N\omega_\y t\right].
\label{eq:Ilin}
\end{align}
Using \eqref{eq:I_N} and \eqref{eq:Ilin}, the information loss of this approximation $\tilde{Y}_t$ in relation to the original
$Y_t$ process is given by
\begin{equation}
\mathcal{I}_N-\mathcal{I}^\mathrm{lin}_N\approx
\frac{1}{2}\ln(4\alpha^2-4\alpha+2),
\label{eq:dI_lin}
\end{equation}
for $t\to\infty$. In this limit the information of the linearized process satisfies
$\mathcal{I}^\mathrm{lin}_N\approx \tilde{\mathcal I}_N$. Thus the continuous linear noise approximation compared to original system dynamics looses the information about the discontinuous binding events,
which can be explained as follows. The linear noise approximation
correctly describes the dynamical behavior of the mean $\avg{y_t}$ and the second order covariance $\avg{y_ty_{t'}}$, i.e., it captures the persistence time $\hat{\tau}_1=\int_0^ty_{t'}\dd t'$ up to its second moment. Since the long time limit renders $\hat{\tau}_1$ Gaussian, the linear noise approximation contains
the information from $\hat{\tau}_1$.
Hence, $\varDelta \mathcal I$, defined in \eqref{eq:def_dI}, is precisely the amount of information that is lost by the continuous linear noise approximation.
This insight constitutes our second main result: The asymmetry parameter $\alpha$
determines whether the linear noise approximation does loose information on the trajectory level or not. Specifically, for $\alpha=1/2$, as it applies to the model from Fig. \ref{fig:4plots}(d), there is no information loss, which makes the linear noise approximation accurate on the trajectory level. For all ``natural'' choices $0\le\alpha\le1$, the lost information is bounded by 1/2\,bit per
measurement, irrespective of the total number of sensors $N$.

\section{Conclusion}
\label{sec:conclusion}
We have shown that the asymmetry parameter, defined in \eqref{eq:def_alpha},
plays a crucial role for the information acquisition of binary sensory networks, since
it determines whether the number of jump events contain additional information about a signal.
Specifically, for a symmetric weight $(\alpha=1/2)$, the
number of jump events do not provide additional information about an external signal $x$, whereas under ``natural conditions'' ($0\le\alpha\le1$) the number of jump events contain up to 1/2\,bit.
Most remarkably, this additional information in the number of jump events
is independent of the number of copies $N$ of binary sensors sensors measuring the same signal $x$.

Furthermore, we have shown that the linear noise approximation, a continuous Brownian motion useful for systems with large copy number ($N\gg1$), describes the probability distribution and linear correlations of the original system up to second order in persistence time. However, discontinuous (and nonlinear) jump events are not covered by this continuous linear noise approximation. Consequently, it looses information compared to the original system that is precisely the additional information from the number of jump events.
Hence, the asymmetry parameter $\alpha$ also determines whether the linear noise approximation looses information about the original system.
Analogously, for $0\le\alpha\le1$ information loss due to the linear noise approximation is less or equal 1/2\,bit, whereas for $\alpha=1/2$ no information is lost, making the
linear noise approximation accurate on the trajectory level.

For future work, it will be interesting to study implications to competing signals \cite{mora15}, where a sensor computes not
just a single signal value. Furthermore, non-equilibrium affinities in multi-state systems may result in non-vanishing currents, such that associated jumps may provide additional information even in the case of a symmetric influence of the signal.
Specifically, it has been found that investing non-equilibrium free energy may reduce the dispersion
of jump related variables \cite{bara15}, see also \cite{bara15a,piet16,ging16}.
Moreover, it will be interesting to study the implications for sensory devices measuring continuously changing signal ramps \cite{mora10,aqui14,hart16},
which requires more than a single measurement discussed here.

Finally, an experimental realization of, e.g., a single electron transistor \cite{kosk14,kosk14a,kosk15},
may be a promising non-biological device to study the information gain $\varDelta\mathcal{I}$ contained in the number of jump events.

\begin{acknowledgments}
 We thank A.C. Barato for fruitful interactions and S. Goldt for useful discussions.
\end{acknowledgments}

\appendix

\section{Information versus error}
\label{sec:information_vs_error}
In this appendix, we briefly review important aspects of differential entropy and mutual information from \cite{cove06}.
Assume we have a stochastic signal $x$ with probability density $P(x)$. Moreover, $m$ is some measurement that is correlated with
$x$ resulting in a conditional distribution $P(m|x)$. Specifically, one can identify $m$ with the sensor state $y$ at time $t$ or with the complete path
$\{y_{t'}\}_{0\le t'\le t}$.
The conditional entropy that quantifies the uncertainty of the signal $x$ given the measurement $m$ is defined as
\begin{align}
 H[x|m]&\equiv-\bavg{\ln P(x|m)}\nonumber\\
 &=-\int\dd x\int \dd m P(m|x)P(x)\ln P(x|m),
 \label{eq:A_Hcond}
\end{align}
where we have used Bayes' formula to get $P(x|m)=P(m|x)P(x)/P(m)$, where $P(m)=\int\dd x' P(m|x')P(x')$. Note that $\int \dd m\cdot $ is a path integral
if one identifies $m$ with the path $\{y_{t'}\}_{0\le t'\le t}$.
It can be shown that any estimator $\hat{x}(m)$ that is only a function of $m$ (i.e. that  uses only the knowledge of $m$) 
trying to optimally estimate $x$
satisfies
the inequality \cite{cove06}
\begin{align}
\mathcal{E}_{\x|\m}^2&\equiv \bbavg{[x-\hat{x}(m)]^2}\nonumber\\&
=\int\dd x\int\dd m [x-\hat{x}(m)]^2P(m|x)P(x)\nonumber\\&
\ge \frac{1}{2\pi\e}\e^{2H[x|m]}
\label{eq:A_error_Hcond}
\end{align}
which can saturate only if $x$ and $m$ are jointly normal distributed.

The uncertainty of the signal before (or without) the measurement is
\begin{equation}
 H[x]=-\bavg{\ln P(x)}=-\int\dd x P(x)\ln P(x).
 \label{eq:A_Hx}
\end{equation}
Specifically, in the main text we have used a normal distributed signal $x$ with zero mean $\avg{x}=0$ and variance $\mathcal{E}_\x^2$. Inserting the normal distribution into \eqref{eq:A_Hx} yields
\begin{equation}
H[x]=\frac{1}{2}\ln(2\pi\e \,\mathcal{E}_\x^2).
\label{eq:A_H}
\end{equation}
Since side information $m$ always reduces the uncertainty, i.e., $H[x|m]\le H[x]$
one can define the non-negative mutual information \cite{cove06}
\begin{equation}
 I[x{:}m]\equiv H[x]-H[x|m]=\bavg{\ln\frac{P(x|m)}{P(x)}},
\label{eq:A_def_Ixm}
\end{equation}
which is the reduction of uncertainty of the signal $x$ due to the measurement $m$.
Using \eqref{eq:A_error_Hcond}, \eqref{eq:A_H} and \eqref{eq:A_def_Ixm}
one obtains the relation between error and information
\begin{equation}
 \frac{\mathcal{E}_{\x|\m}^2}{\mathcal{E}_\x^2}\ge\e^{-2 I[x{:}m]}.
\end{equation}
Note that this inequality saturates if $m|x$ is normally distributed.

The mutual information is symmetric $I[x{:}m]=I[m{:}x]$. Moreover, for
jointly Gaussian distribution of $\bz^\T\equiv(x,\boldsymbol{m})$ with covariance
\begin{equation}
 \bSigma\equiv\avg{\bz\bz^\T}-\avg{\bz}\avg{\bz}^\T\equiv
 \begin{pmatrix}
  \mathcal{E}_\x^2&\bbb^\T\\
  \bbb&\bSigma_\m
 \end{pmatrix}
 \label{eq:A_SigmaZ}
\end{equation}
the mutual information reads
\begin{equation}
 I[x{:}\boldsymbol{m}]=\frac{1}{2}\ln\frac{\mathcal{E}_{\x}^2}{\mathcal{E}_{\x|\m}^2}
 =\frac{1}{2}\ln\frac{\det(\bSigma_\m)}{\det(\bSigma_{\m|x})},
 \label{eq:A_Igauss}
\end{equation}
where the conditional variances can be calculated with the Schur complements of \eqref{eq:A_SigmaZ}, see \cite{cott74},
\begin{align}
 \mathcal{E}_{\x|\m}^2&=\mathcal{E}_\x^2-\bbb^\T\bSigma_\m^{-1}\bbb,
 \label{eq:A_schur1}\\
 \bSigma_{\m|\x}&=\bSigma_\m-\bbb\frac{1}{\mathcal{E}_\x^2}\bbb^\T.
 \label{eq:A_schur2}
\end{align}
It will convenient to express the mutual information in terms of
$\bSigma_{\m|\x},\mathcal{E}_\x^2$ and $\bbb$. Therefore, the mutual information between jointly Gaussian variables reads
\begin{align}
 I[x{:}\boldsymbol{m}]&=\frac{1}{2}\ln\left[\frac{\det(\bSigma_{\m|\x}+\bbb\frac{1}{\mathcal{E}_\x^2}\bbb^\T)}{\det(\bSigma_{\m|\x})}\right]\nonumber\\
&= \frac{1}{2}\ln\left[\det\left(\boldsymbol{1}+\bSigma_{\m|\x}^{-1}\bbb\frac{1}{\mathcal{E}_\x^2}\bbb^\T\right)\right]\nonumber\\
&=\frac{1}{2}\ln\left(1+\frac{\bbb^\T\bSigma_{\m|\x}^{-1}\bbb}{\mathcal{E}_\x^2}\right).
\label{eq:A_Iformula}
\end{align}
In the first step we have used \eqref{eq:A_Igauss} and \eqref{eq:A_schur2}.
According to the matrix determinant lemma, the determinant of the identity matrix plus a dyadic product can be written as scalar product, which has been used in the final step.

\section{Generating function}
\label{sec:gen_func}
In this appendix, we will define the generating function $\mathcal{G}(\bs,t)$, which is related to
the conditional path weight $P(\{y_{t'}\}_{0\le t'\le t}|x)$ defined in Eq. \eqref{eq:def_pathprob} of the main text.
We will calculate conditional expectation values $\avg{\cdots}_x$ that will all depend on the signal value $x$
and satisfy $\avg{\cdots}=\int\dd xP(x)\avg{\cdots}_x$.

\subsection{Master equation and generating function}
The binary system with states $y=0,1$, which evolves stochasticly in time, satisfy the master equation
\begin{equation}
 \begin{aligned}
  \frac{\del}{\del t}P_t(0)&=w_{10}(x)P_t(1)-w_{01}(x)P_t(0),\\
  \frac{\del}{\del t}P_t(1)&=w_{01}(x)P_t(0)-w_{10}(x)P_t(1),
 \end{aligned}
\label{eq:A_master}
\end{equation}
where $P_t(y)$ is the probability for being in state $y$ at time $t$. Note that $P_t(1)=\avg{y_t}_x$ holds. The master equation can be written in matrix notation
\begin{equation}
\frac{\dd }{\dd t}\boldsymbol{P}_t=\mat{L}\boldsymbol{P}_t,
\label{eq:A_master_matrix}
\end{equation}
where
\begin{equation}
\mat{L}\equiv
\begin{pmatrix}
	-w_{01}(x)&w_{10}(x)\\
	w_{01}(x)&-w_{10}(x)
\end{pmatrix}\quad\text{and}\quad
\boldsymbol{P}_t\equiv
\begin{pmatrix}
	P_t(0)\\
	P_t(1)
\end{pmatrix}.
\label{eq:A_def_matrix}
\end{equation}
For later convenience, we have dropped notationally the explicit dependence of the generator $\mat{L}$
on $x$.
With initial distribution $\boldsymbol{P}_0$ the formal time dependent solution reads
$\boldsymbol{P}_t=\exp(\mat{L}t)\boldsymbol{P}_0$, which can be associated with
the path weight
\begin{equation}
P(\{y_{t'}\}_{0\le t'\le t}|x)=P_0(y)(L_{01})^{\hat{n}_{10}}(L_{10})^{\hat{n}_{01}}\e^{L_{00}\hat{\tau}_0+L_{11}\hat{\tau}_1},
\label{eq:A_path weight_L}
\end{equation}
see Eq. \eqref{eq:def_pathprob} in the main text.
Specifically, $(\boldsymbol{P}_t)_y$ can be obtained from \eqref{eq:A_path weight_L} by summing over all paths $\{y_{t'}\}_{0\le t'\le t}$ with $y_t=y$.

The generating function is defined as
\begin{equation}
 \mathcal{G}(\bs,t)\equiv\boldsymbol{\mathcal{V}}^\T\exp\left[\tilde{\boldsymbol{\mathcal{L}}}(\bs)t\right]\boldsymbol{P}_0,
\label{eq:A_Gs}
\end{equation}
where  $\boldsymbol{\mathcal{V}}^\T\equiv(1,1)$ and the modified generator reads  $\tilde{\boldsymbol{\mathcal{L}}}(\bs)_{yy'}\equiv L_{yy'}s_{y'y}$, i.e.,
\begin{equation}
\tilde{\boldsymbol{\mathcal{L}}}(\bs)=
\begin{pmatrix}
	-w_{01}(x)s_{00}&w_{10}(x)s_{10}\\
	w_{01}(x)s_{01}&-w_{10}(x)s_{11}
\end{pmatrix}.
\label{eq:A_mod_gen}
\end{equation}
Note that $\boldsymbol{\mathcal{V}}^\T\boldsymbol{P}_t=1$ and $\boldsymbol{\mathcal{V}}^\T\mat{L}=0$.

From now on ``$\bs=1$'' means $s_{yy'}=1$ for all $y,y'$. Specifically, $\tilde{\boldsymbol{\mathcal{L}}}(\bs)\big|_{\boldsymbol{s}=1}=\mat{L}$. The aim of introducing this generating function 
\eqref{eq:A_Gs} is to calculate moments of $\hat{n}_{yy'}$ and $\hat{\tau}_y$, which are functionals of the path $\{y_{t'}\}_{0\le t'\le t}$.

\subsection{Moments and conditional covariance}
Since one can interpret \eqref{eq:A_Gs} as the
path integral over the path weight \eqref{eq:A_path weight_L} with replaced generator $L_{yy'}\to s_{y'y}L_{yy'}$, we are able to derive the formulas for the first moments
by comparing Eqs. \eqref{eq:A_def_matrix} -- \eqref{eq:A_mod_gen},
which read
\begin{equation}
 \avg{\hat{\tau}_y}_x=\frac{1}{L_{yy}}\left.\frac{\del\mathcal{G}}{\del s_{yy}}\right|_{\bs=1}
 \label{eq:A_moments_1st_tau}
\end{equation}
and
\begin{equation}
 \avg{\hat{n}_{yy'}}_x=\left.\frac{\del\mathcal{G}}{\del s_{y'y}}\right|_{\bs=1}.
 \label{eq:A_moments_1st_jump}
\end{equation}
Similarly, the second moments are given by
\begin{align}
 \avg{{\hat{\tau}_y}^2}_x&=\frac{1}{L_{yy}^2}\left.\frac{\del^2\mathcal{G}}{\del s_{yy}^2}\right|_{\bs=1},
 \label{eq:A_moments_2nd_tau}\\
 \avg{\hat{n}_{yy'}\hat{\tau}_{\tilde{y} }}_x&=\frac{1}{L_{\tilde{y}\tilde{y}}}\left.\frac{\del^2\mathcal{G}}{\del s_{y'y}\del s_{\tilde{y}\tilde{y}}}\right|_{\bs=1},\label{eq:A_moments_1st_taujump}\\
 \avg{\hat{n}_{yy'}^2}_x&=\left.\frac{\del^2\mathcal{G}}{\del s_{y'y}^2}\right|_{\bs=1}+\left.\frac{\del\mathcal{G}}{\del s_{y'y}}\right|_{\bs=1}.
 \label{eq:A_moments_2nd_jump}
\end{align}
In the limit $t\gg1/[w_{01}+w_{10}]$, the maximum eigenvalue (maximum real part) of the modified generator
\eqref{eq:A_mod_gen} dominates, which is
\begin{multline}
 \lambda_\mathrm{max}(\bs)=
-\frac{w_{01}(x)s_{00}+w_{10}(x)s_{11}}{2}\\
+\sqrt{w_{01}(x)w_{10}(x)s_{01}s_{10}+\frac{[w_{01}(x)s_{00}-w_{10}(x)s_{11}]^2}{4}}.
\label{eq:A_eig}
\end{multline}
In this limit, the first moments become
\begin{align}
  \avg{\hat{\tau}_1}_x&\approx
\frac{w_{01}(x)t}{w_{01}(x)+w_{01}(x)}\label{eq:A_moment1a}\\
\avg{\hat{n}_\mathrm{tot}}_x&\equiv\avg{\hat{n}_{01}}_x+\avg{\hat{n}_{10}}_x \approx
\frac{2w_{01}(x)w_{10}(x)t}{w_{01}(x)+w_{01}(x)}.
\label{eq:A_moment1b}
\end{align}
Analogously, the signal dependent conditional covariance matrix, where $m=(\hat{\tau}_1,\hat{n}_\mathrm{tot})$,
is defined as
\begin{multline}
 \boldsymbol{\Sigma}_{\m|\x}(x)\equiv\\
 \begin{pmatrix}
  \avg{\hat{\tau}_1^2}_x-\avg{\hat{\tau}_1}_x^2&\avg{\hat{\tau}_1\hat{n}_\mathrm{tot}}_x
  -\avg{\hat{\tau}_1}_x\avg{\hat{n}_\mathrm{tot}}_x\\
  \avg{\hat{\tau}_1\hat{n}_\mathrm{tot}}_x
  -\avg{\hat{\tau}_1}_x\avg{\hat{n}_\mathrm{tot}}_x&\avg{\hat{n}_\mathrm{tot}^2}_x-\avg{\hat{n}_\mathrm{tot}}_x^2
 \end{pmatrix},
\end{multline}
which, for $t\gg1/(w_{01}+w_{10})$, reads
\begin{multline}
 \boldsymbol{\Sigma}_{\m|\x}(x)\approx
\frac{2w_{01}(x)w_{10}(x)t}{[w_{10}(x)+w_{01}(x)]^3}\\\times
 \begin{pmatrix}
  1&
 w_{10}(x)-w_{01}(x)\\
 w_{10}(x)-w_{01}(x)&2[w_{01}(x)^2+w_{10}(x)^2]
 \end{pmatrix}.
 \label{eq:A_Sigma_m|x}
\end{multline}
For weak signals ($\mathcal{E}_\x^2\ll 1$), the conditional variance can then be obtained by
\begin{equation}
 \boldsymbol{\Sigma}_{\m|\x}\equiv\int \dd x P(x)\boldsymbol{\Sigma}_{\m|\x}(x)\approx
 \boldsymbol{\Sigma}_{\m|\x}(0)+\boldsymbol{\Sigma}_{\m|\x}''(0)\mathcal{E}_\x^2,
\end{equation}
where for our purposes $\boldsymbol{\Sigma}_{\m|\x}\approx \boldsymbol{\Sigma}_{\m|\x}(0)$ suffices for Eq. \eqref{eq:A_Iformula}.
Note that we show in Appendix \ref{sec:method} how one can perform the calculations for more complicated systems with more than two states, where it maybe difficult to
determine the eigenvalue \eqref{eq:A_eig}.

\subsection{Mutual information from generating function}
In the following two subsections, we derive the weak signal approximation
$\mathcal{E}_\x^2\ll1$ for the mutual information used in the main text, i.e., each approximation ``$\approx$'' will be exact up to order $\mathcal{E}_\x^4$.
All our approximations are fairly accurate for $\mathcal{E}_\x=0.3$ as used in the main text, which corresponds to a ``concentration change'' of about 30\,\%.
In Appendix \ref{sec:robust} we will show that higher order corrections vanish for our main result [Eq. \eqref{eq:def_dI} in the main text] in the long time limit.

For the mutual information \eqref{eq:A_Iformula}, we need $\boldsymbol{\Sigma}_{\m|\x}(0)$, which is given by \eqref{eq:A_Sigma_m|x} and
$\bbb$ from \eqref{eq:A_SigmaZ}, which is
\begin{align}
 \bbb&\equiv
 \begin{pmatrix}
  \avg{x\hat{\tau}_1}-\avg{x}\avg{{\hat{\tau}_1}}\\
  \avg{x\hat{n}_\mathrm{tot}}-\avg{x}\avg{\hat{n}_\mathrm{tot}}
 \end{pmatrix}
 =
  \begin{pmatrix}
  \avg{x\hat{\tau}_1}\\
  \avg{x\hat{n}_\mathrm{tot}}
 \end{pmatrix}\nonumber\\
 &=\int\dd xP(x)
   \begin{pmatrix}
  x\avg{\hat{\tau}_1}_x\\
  x\avg{\hat{n}_\mathrm{tot}}_x
 \end{pmatrix}\nonumber\\
 &\approx\mathcal{E}_\x^2\frac{\del}{\del x}\left.
 \begin{pmatrix}
 \avg{\hat{\tau}_1}_x\\
  \avg{\hat{n}_\mathrm{tot}}_x
 \end{pmatrix}\right|_{x=0}.
\end{align}
Using the definition of the asymmetry parameter Eq. \eqref{eq:def_alpha} in the main text
and Eqs. \eqref{eq:A_moment1a} and \eqref{eq:A_moment1b},
we obtain
\begin{equation}
 \bbb\approx
  \frac{\mathcal{E}_\x^2w_{01}(0)w_{10}(0)t}{[w_{01}(0)+w_{10}(0)]^2}
 \begin{pmatrix}
1  \\
  2 (1-\alpha)w_{10}(0)-2\alpha w_{01}(0)
 \end{pmatrix}.
 \label{eq:A_b}
\end{equation}

The mutual information between jointly Gaussian variables then reads
\begin{align}
 \mathcal{I}
&
\approx I[x{:}\hat{\tau}_1,\hat{n}_\mathrm{tot}]\approx
\frac{1}{2}\ln\left(1+\frac{\bbb^\T\boldsymbol{\Sigma}_{\m|x}(0)^{-1}\bbb}{\mathcal{E}_\x^2}\right)\nonumber\\
&=\frac{1}{2}\ln\left[1+(2\alpha^2-2\alpha+1)
\frac{\mathcal{E}_\x^2w_{01}(0)w_{10}(0)t}{w_{01}(0)+w_{10}(0)}
\right],
\label{eq:A_Iges}
\end{align}
where we have inserted \eqref{eq:A_Sigma_m|x} and \eqref{eq:A_b} in \eqref{eq:A_Iformula}.

Similarly for $\tilde{\mathcal{I}}$, where $m=\hat{\tau}_1$, one obtains the coarse grained mutual information by using the upper right component of \eqref{eq:A_Sigma_m|x} and the first component of \eqref{eq:A_b}
\begin{align}
\tilde{\mathcal{I}}&\approx I[x{:}\hat{\tau}_1]
\approx
\frac{1}{2}\ln\left(1+\frac{(\bbb_{1})^2}{[\bSigma_{\m|x}(0)]_{11}\mathcal{E}_\x^2}\right)\nonumber\\
&=\frac{1}{2}\ln\left[1+\frac{1}{2}
\frac{\mathcal{E}_\x^2w_{01}(0)w_{10}(0)t}{w_{01}(0)+w_{10}(0)}
\right].
\label{eq:A_Itilde}
\end{align}
Comparing \eqref{eq:A_Iges} and \eqref{eq:A_Itilde} leads immediately to Eq. \eqref{eq:def_dI} in the main text,
for weak signals. We show in Appendix \ref{sec:robust} that higher order corrections
for the difference between \eqref{eq:A_Iges} and \eqref{eq:A_Itilde} cancel.

\subsection{Multiple independent sensors}
\label{sec:multi}
For a sensor array $\by_t=(y^{(1)}_t,\ldots, y^{(1)}_t)$ of $N$ independent sensors, see top panel of Fig. \ref{fig:sampletraj} in the main text for an illustration, the path weight becomes
\begin{equation}
 P(\{\by_{t'}\}_{0\le t'\le t}|x)=\prod_{i=1}^NP(\{y^{(i)}_{t'}\}_{0\le t'\le t}|x),
 \label{eq:A_def_N_pathprob}
\end{equation}
where each path weight $P(\{y^{(i)}_{t'}\}_{0\le t'\le t}|x)$ is given by
Eq. \eqref{eq:def_pathprob} in the main text with respective individual functionals $\hat{n}_{yy'}^{(i)}$ and $\tau_y^{(i)}$ with $\tau_0^{(i)}+\tau_1^{(i)}=t$.
The resulting total path weight is given by Eq. \eqref{eq:def_N_pathprob} in the main text.
The functional of the path $\{\by_{t'}\}_{0\le t'\le t}$
\begin{equation}
\boldsymbol{f}\equiv
 \begin{pmatrix}
 Y_0\\ \tau_1\\\hat{n}_{01}\\\hat{n}_{10}
 \end{pmatrix}
=\sum_{i=1}^N
\begin{pmatrix}
 y_0^{(i)}\\\tau_1^{(i)}\\\hat{n}_{01}^{(i)}\\\hat{n}_{10}^{(i)}
\end{pmatrix}
\equiv\sum_{i=1}^N\boldsymbol{f}^{(i)},
\end{equation}
fully determines the path weight \eqref{eq:A_def_N_pathprob} that is given by Eq. \eqref{eq:def_N_pathprob} in the main text.
Most importantly, the first moments satisfy
\begin{equation}
 \avg{\boldsymbol{f}}_x=\sum_{i=1}^N\avg{\boldsymbol{f}^{(i)}}_x=N \avg{\boldsymbol{f}^{(1)}}_x
 \label{eq:a_ffirst}
\end{equation}
and similarly, the variance is given by
\begin{multline}
 \avg{\boldsymbol{f}\boldsymbol{f}^\T}_x-\avg{\boldsymbol{f}}_x\avg{\boldsymbol{f}^\T}_x\\=\sum_{i,j}\left[\avg{\boldsymbol{f}^{(i)}\boldsymbol{f}^{(j)}{}^\T}_x-\avg{\boldsymbol{f}^{(i)}}_x\avg{\boldsymbol{f}^{(j)}{}^\T}_x\right]\\
 =N\left[\avg{\boldsymbol{f}^{(1)}\boldsymbol{f}^{(1)}{}^\T}_x-\avg{\boldsymbol{f}^{(1)}}_x\avg{\boldsymbol{f}^{(1)}{}^\T}_x\right],
  \label{eq:a_fsecond}
\end{multline}
where we have used that $\avg{\boldsymbol{f}^{(i)}\boldsymbol{f}^{(j)}}_x=\avg{\boldsymbol{f}^{(i)}}_x\avg{\boldsymbol{f}^{(j)}}_x$ for $i\neq j$ and that each moment of $\boldsymbol{f}^{(i)}$
coincides with the respective moment of $\boldsymbol{f}^{(1)}$.

Using similar arguments as for the single sensor case one can see that each forward jump must be compensated by a reverse jump, i.e.,
the coarse grained statistics
$\boldsymbol{m}\equiv(\hat{\tau}_1,\hat{n}_\mathrm{tot})=(\hat{\tau}_1,\hat{n}_{01}+\hat{n}_{10})$
satisfies
\begin{equation}
 \mathcal{I}_N=I[x:\boldsymbol{f}]\approx I[x:\boldsymbol{m}].
\end{equation}
First, from \eqref{eq:a_ffirst} it follows $\bbb_N=N\bbb$, where $\bbb$ is given from the single sensor solution
\eqref{eq:A_b}. Second, from \eqref{eq:a_fsecond} it follows that $\bSigma_{\m|\x,N}(x)=N\bSigma_{\m|\x}(x)$. Substituting $\bbb\to\bbb_N$ and $\bSigma_{\m|\x}(0)\to \bSigma_{\m|\x,N}(0)$ in \eqref{eq:A_Iges}
yields the mutual information
\begin{align}
 \mathcal{I}_N
&
\approx
\frac{1}{2}\ln\left(1+N\frac{\bbb^\T\boldsymbol{\Sigma}_{\m|x}(0)^{-1}\bbb}{\mathcal{E}_\x^2}\right)\nonumber\\
&=\frac{1}{2}\ln\left[1+N(2\alpha^2-2\alpha+1)
\frac{\mathcal{E}_\x^2w_{01}(0)w_{10}(0)t}{w_{01}(0)+w_{10}(0)}
\right]
\label{eq:A_Iges_N}
\end{align}
between the time series $\{\by_{t'}\}_{0\le t'\le t}$ of an array of $N$ sensors and the signal $x$ for $t\gg 1/\omega_\y$,
which is precisely equation \eqref{eq:I_N} in the main text, where $\omega_\y\equiv w_{01}(0)+w_{01}(0)$ and $p\equiv w_{01}(0)/[w_{01}(0)+w_{10}(0)]$.

Similarly, one obtains
 \begin{equation}
 \tilde{\mathcal{I}}_N\equiv I[x:\hat{\tau}_1]
\approx
\frac{1}{2}\ln\left[1+\frac{N}{2}
\frac{\mathcal{E}_\x^2w_{01}(0)w_{10}(0)t}{w_{01}(0)+w_{10}(0)}
\right].
\label{eq:A_Itilde_N}
\end{equation}

Comparing \eqref{eq:A_Iges_N} and \eqref{eq:A_Itilde_N} results in
\begin{equation}
\lim_{t\to\infty}\left(\mathcal{I}_N-\tilde{\mathcal{I}}_N\right)=\frac{1}{2}\ln\left(4\alpha^2-4\alpha+2\right),
\end{equation}
which does not depend on $N$ and is precisely the loss of information
$\varDelta \mathcal{I}$ of a single sensor that is given by
\eqref{eq:def_dI} in the main text.

\section{Robustness against non-linearities}
\label{sec:robust}
In this appendix, we show that our main result is still valid if $x$
is broadly distributed with probability $P_\x(x)$. We derive an approximation $\mathcal{I}^\infty$ similar
to \eqref{eq:A_Iges} [Eq. \eqref{eq:I_N} in the main text] that converges more slowly to $\mathcal{I}$ but can be used for strong signals (e.g., $\mathcal{E}_\x^2>1)$.

We discuss $\boldsymbol{m}\equiv (m_1,\ldots,m_d)$
for $d=2$ with $m_1\equiv\hat{\tau}_1$ and $m_2\equiv\hat{n}_\text{tot}$.
The mean $\bmu(x)\equiv\avg{\boldsymbol{m}}_x$
is determined by \eqref{eq:A_moment1a} and \eqref{eq:A_moment1b}.
The covariance matrix \eqref{eq:A_Sigma_m|x} then determines
the
conditional probability, which reads (here $d=2$)
\begin{equation}
 P(\boldsymbol{m}|x)=\frac{\e^{l(\boldsymbol{m}|x)}}{\sqrt{(2\pi)^d\det[\bSigma_{\m|\x}(x)]}},
 \label{eq:A_Pmx_last}
\end{equation}
where
\begin{equation}
 l(\boldsymbol{m}|x)\equiv-\frac{1}{2}[\boldsymbol{m}-\bmu(x)]^\T\bSigma_{\m|\x}(x)^{-1}[\boldsymbol{m}-\bmu(x)].
 \label{eq:A_lmx_last}
\end{equation}
Since $\bSigma_{\m|\x}(x),\bmu(x)\sim t$ holds, we obtain
a substantial weight only for $\boldsymbol{m}\approx\bmu(x)$ in the limit $t\to\infty$.

\begin{figure}
 \centering
 \includegraphics{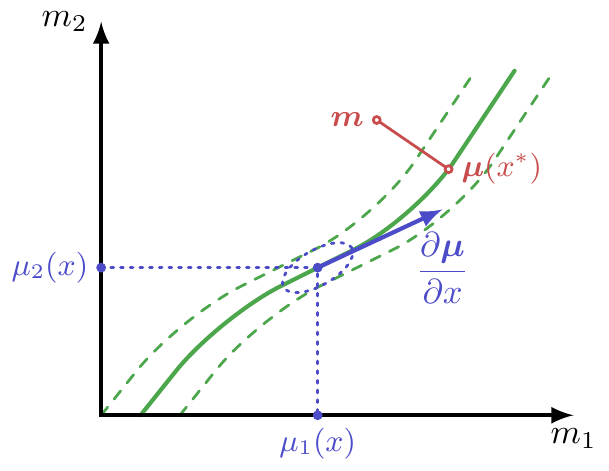}
 \caption{Schematic view of marginal distribution $\boldsymbol{m}$ (green dashed lines) and conditional distribution
 $\boldsymbol{m}|x$ (blue dotted ellipse).}
 \label{fig:A_trace}
\end{figure}

We will now use Fig. \ref{fig:A_trace} as a blue print to derive an approximation for the marginal distribution of $\boldsymbol{m}$,
which can be used to determine the mutual information between $\boldsymbol{m}$ and $x$ in the long time limit exactly. The green solid curve corresponds
to the one dimensional line $\boldsymbol{m}=\bmu(x)$ where $x\in\mathbb{R}$.
The dashed line illustrates the width of the marginal distribution of $\boldsymbol{m}$. The blue point with the dotted ellipse corresponds
to a single point $\bmu(x)=(\mu_1(x),\mu_2(x))^\T$, where the ellipse emphasizes
the width $\bSigma_{\m|\x}(x)$ of the conditional distribution \eqref{eq:A_Pmx_last}, where $x$ is given.
It is important to note that an increase of variance and mean that increases linearly in time implies $\boldsymbol{m}-\bmu(x)\sim\sqrt{t}$ and $\del_x\bmu(x)\sim t$, leading to
\begin{equation}
\frac{\del^2}{\del x^2} l(\boldsymbol{m}|x)\approx-\frac{\del\bmu(x)}{\del x}^\T\bSigma_{\m|\x}(x)^{-1}\frac{\del\bmu(x)}{\del x}.
 \label{eq:A_d2lmx_last}
\end{equation}

For the marginal path weight one has to integrate \eqref{eq:A_Pmx_last} over $x$. 
Using saddle point approximation we obtain for the path weight
\begin{align}
 P_\m(\boldsymbol{m})&\equiv\int\dd x P_{\m|\x}(\boldsymbol{m}|x)P_\x(x)\nonumber\\
 &\approx\frac{P_\x(x^*(\boldsymbol{m}))\e^{l(\boldsymbol{m}|x^*)}}{\sqrt{(2\pi)^{d-1}\det[\bSigma_{\m|\x}(x^*)]\frac{\del\bmu}{\del x}^\T\bSigma_{\m|\x}(x^*)^{-1}\frac{\del\bmu}{\del x}}},
 \label{eq:A_Pm_last}
\end{align}
where $x^*\equiv x^*(\boldsymbol{m})$ maximizes \eqref{eq:A_lmx_last} for a given $\boldsymbol{m}$,
as indicated with two points connected via a red solid line in Fig. \ref{fig:A_trace}.
\begin{figure}
 \centering
 \includegraphics{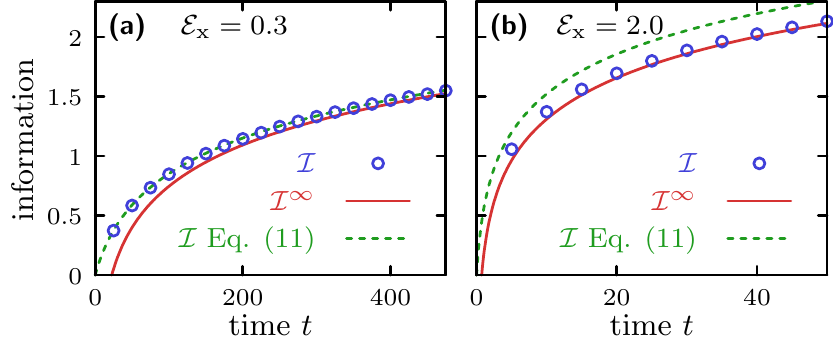}
 \caption{Nonlinear approximation $\mathcal{I}^\infty$ versus exact result with asymmetric rates ($\alpha=0$). The numerical solution of the exact mutual information $\mathcal{I}$ is indicated as blue open circles. The red solid line indicates the approximation from Eq. \eqref{eq:I_N} of the main text, which is here equivalent to \eqref{eq:A_Iges}, and assumes a narrow-width Gaussian distribution of $x$. Parameter:
 $w_{01}=\e^x,w_{10}=1$, $\alpha=0$. (a) Signal standard deviation $\mathcal{E}_\x=0.3$  [same as in Fig. \ref{fig:4plots}(a) in the main text]. (b) $\mathcal{E}_\x=2$.}
 \label{fig:0.3vs2.0}
\end{figure}

We can approximate the mutual information between $x$ and $\boldsymbol{m}$,
\begin{align}
 I[x{:}\boldsymbol{m}]&\equiv\int \dd \boldsymbol{m}\int\dd x P_\x(x)P_{\m|\x}(\boldsymbol{m}|x)\ln\frac{P_{\m|\x}(\boldsymbol{m}|x)}{P_{\m}(\boldsymbol{m})},
\end{align}
by using Eqs. \eqref{eq:A_Pmx_last}, \eqref{eq:A_Pm_last} and the relation
\begin{equation}
 l(\boldsymbol{m}|x)-l(\boldsymbol{m}|x^*)=\frac{1}{2}\frac{\del l(\boldsymbol{m}|x^*)}{\del x^2}(x-x^*)^2+\mathrm{O}(x-x*)^3,
 \label{eq:Adl}
\end{equation}
which becomes $-1/2=\ln\sqrt{1/\e}$ after averaging over $x,\boldsymbol{m}$.
Finally, we set $x^*=x$ and find
\begin{align}
 I[x{:}\boldsymbol{m}]
 &\approx\int \dd x P_\x(x)\ln\Bigg[\frac{1}{P_\x(x)}\sqrt{\frac{\frac{\del\bmu}{\del x}^\T\bSigma_{\m|\x}(x)^{-1}\frac{\del\bmu}{\del x}}{2\pi \e}}\Bigg]
 \nonumber\\
 &\equiv\mathcal{I}^\infty.
 \label{eq:A_Iinfty}
\end{align}
We note that Eq. \eqref{eq:A_Iinfty}
becomes exact in the long time limit ($t\to\infty$).

With Fig. \ref{fig:0.3vs2.0} we confirm that \eqref{eq:A_Iinfty} is indeed accurate
for sufficient long time. For strong signals ($\mathcal{E}_\x=2$) the approximation from
Eq. \eqref{eq:A_Iges}, which is Eq. \eqref{eq:I_N} in the main text, fails. In this case one has to use
\eqref{eq:A_Iinfty} instead. Moreover, one can see that for a narrow-width Gaussian the approximation from Eq. \eqref{eq:I_N} in the main text is quite accurate even for ``finite time''.
Note that the parameters from Fig. \ref{fig:0.3vs2.0}(a) are the same as in Fig. \ref{fig:4plots}(a)
in the main text.

\begin{figure}
 \centering
 \includegraphics{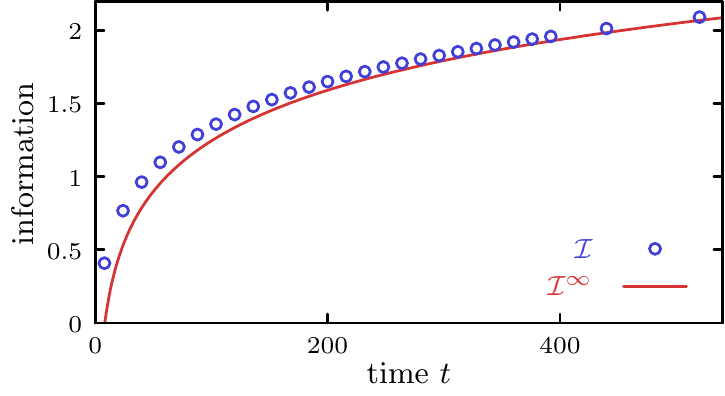}
 \caption{Nonlinear approximation $\mathcal{I}^\infty$ versus exact result with asymmetric rates ($\alpha=0$). The signal is uniformly distributed within $-1\le x\le 1$. Parameter:
 $w_{01}=\e^x,w_{10}=1$, $\alpha=0$.}
 \label{fig:rectangle}
\end{figure}

This approximation can also be used for non-Gaussian distributed signals.
One example of a uniformly distributed signal is illustrated in Fig. \ref{fig:rectangle}, where $P_\x(x)=1/2$ is used for $-1\le x\le 1$.
For $t\to\infty$, the approximation $\mathcal{I}^\infty$ approaches $\mathcal{I}$. The convergence is more slowly than for a Gaussian distributed signal with the same width.

For $d=1$ and $m_1=\hat{\tau}_1$,
the coarse-grained version of \eqref{eq:A_Iinfty}
reads
\begin{align}
 I[x{:}m_1]&\approx\int \dd x P_\x(x)\ln\Bigg[\frac{1}{P_\x(x)}\sqrt{\frac{\left|\frac{\del \mu_1(x)}{\del x}\right|^2}{ 2\pi\e [\bSigma_{\m|\x}(x)]_{11}}}\Bigg]
 \nonumber\\
 &\equiv\tilde{\mathcal{I}}^\infty,
 \label{eq:A_Icginfty}
\end{align}
with equality for $t\to\infty$.
Note that from \eqref{eq:A_Sigma_m|x} follows $[\bSigma_{\m|\x}(x)]_{11}=2w_{01}(x)w_{10}(x)t/[w_{01}(x)+w_{10}(x)]^3$.

The difference between \eqref{eq:A_Iinfty} and \eqref{eq:A_Icginfty} yields an information gain of
the jump events
\begin{multline}
\varDelta \mathcal{I}^\infty\equiv\mathcal{I}^\infty-\tilde{\mathcal{I}}^\infty\\
=
 \int\dd xP_\x(x)\frac{1}{2}\ln \left[\frac{[\bSigma_{\m|\x}(x)]_{11}\frac{\del\bmu}{\del x}^\T\bSigma_{\m|\x}(x)^{-1}\frac{\del\bmu}{\del x}}{\left|\frac{\del \mu_1(x)}{\del x}\right|^2}\right].
 \label{eq:A_IinfDiff}
\end{multline}
The same calculation as in \eqref{eq:A_b} by using \eqref{eq:A_moment1a}, \eqref{eq:A_moment1b} and Eq. \eqref{eq:def_alpha} from the main text
can be used to obtain
\begin{multline}
 \frac{\del \bmu}{\del x}
 =\frac{\del}{\del x}
 \begin{pmatrix}
  \mu_1(x)\\
  \mu_2(x)
 \end{pmatrix}
 =\frac{\del}{\del x}
 \begin{pmatrix}
  \avg{\hat{\tau}_1}_x\\
  \avg{\hat{n}_\text{tot}}_x
 \end{pmatrix}\\
 \approx
 \frac{w_{01}(x)w_{10}(x)t}{[w_{01}(x)+w_{10}(x)]^2}
 \begin{pmatrix}
1  \\
  2 (1-\alpha)w_{10}(x)-2\alpha w_{01}(x)
 \end{pmatrix},
\end{multline}
which together with \eqref{eq:A_Sigma_m|x} finally yields
\begin{equation}
 \varDelta \mathcal{I}^\infty=
 \int\dd xP_\x(x)\frac{1}{2}\ln(4\alpha^2-4\alpha+2).
\end{equation}
This result implies Eq. \eqref{eq:def_dI} from the main text, since the approximations \eqref{eq:A_Icginfty} and \eqref{eq:A_Iinfty} are accurate for $t\to\infty$.

Finally, the derivation from Appendix \ref{sec:multi} can be adopted to an array of $N$ independent sensors. Since
mean and covariance is proportional to $N$, the expression \eqref{eq:A_IinfDiff} does not depend on the number of sensors.

\section{Method of References \cite{bara15a,wach15,koza99} for joint dispersions}
\label{sec:method}
In the limit $t\to\infty$, the generating function \eqref{eq:A_Gs}
has the form $\mathcal{G}(\bs,t)\approx\phi(\bs)\e^{\lambda_\mathrm{max}(\bs)t}$ with equality for $t\to\infty$. Moreover, the conservation of probability determines $\phi(\bs)|_{\bs=1 }=1$ and
the existence of a steady state requires
$\lambda_\mathrm{max}(\bs)|_{\bs =1}=0$. Let $a=s_{ij}$ and $b=s_{kl}$.
Thus one obtains
\begin{align}
 \left.\frac{\del\mathcal{G}(\bs,t)}{\del a}\right|_{\bs=1}&=\left.\frac{\del \lambda_\mathrm{max}(\bs)}{\del a}\right|_{\bs=1} t+\left.\frac{\del \phi(\bs)}{\del a}\right|_{\bs=1}\nonumber\\
 &\approx
 \left.\frac{\del \lambda_\mathrm{max}(\bs)}{\del a}\right|_{\bs=1} t.
 \label{eq_A_1st_derivative}
\end{align}
Similarly, the dispersion is related to the second derivative
\begin{multline}
  \left.\frac{\del^2\mathcal{G}(\bs,t)}{\del a\del b}\right|_{\bs=1}-\left.\frac{\del\mathcal{G}(\bs,t)}{\del a}\right|_{\bs=1}\left.\frac{\del\mathcal{G}(\bs,t)}{\del b}\right|_{\bs=1}\\
  \left.\approx\frac{\del^2\lambda_\mathrm{max}(\bs)}{\del a\del b}\right|_{\bs=1}t.
  \label{eq_A_2nd_derivative}
\end{multline}

Following the same procedure as in \cite{koza99} (see also \cite{bara15a,wach15}) one can obtain the derivatives of the maximal eigenvalue directly from the characteristic polynomial
\begin{multline}
 \chi_{\tilde{\mathcal{L}}(\bs)}(\lambda)\equiv \det[\tilde{\mathcal{L}}(\bs)-\lambda\mathbf{1}]=\prod_i(\lambda_i(\bs)-\lambda)
\\
 =\overbrace{\prod_i\lambda_i(\bs)}^{\equiv C_0(\bs)}
-\lambda\overbrace{\sum_{j}\prod_{j\neq i}\lambda_i(\bs)}^{\equiv C_1(\bs)}
 +\lambda^2\overbrace{\sum_{i< j}\prod_{k\neq i,j}\lambda_k(\bs)}^{\equiv C_2(\bs)}+\mathrm{O}(\lambda)^3, \hidewidth\\
\end{multline}
where $\lambda_i(\bs)$ are the eigenvalues of $\tilde{\mathcal{L}}(\bs)$, which need not to be known explicitly.
Since $i\neq \mathrm{max}$ implies $\lambda_i(\bs)|_{\bs=1}\neq 0$ and $\lambda_\mathrm{max}(\bs)|_{\bs=1}=0$, we find
\begin{align}
 C_0(\bs)|_{\bs=1}&=0,\\
 C_1(\bs)|_{\bs=1}&=\prod_{i\neq \mathrm{max}}\lambda_{i}(\bs)|_{\bs=1},\\
 C_2(\bs)|_{\bs=1}&=\sum_{i\neq \mathrm{max}}\left[\prod_{j\neq i, \mathrm{max}}\lambda_{j}(\bs)|_{\bs=1}\right].
\end{align}
Hence, one can identify
\begin{equation}
 \left.\frac{\del \lambda_\mathrm{max}(\bs)}{\del a}\right|_{\bs=1}=\left[\frac{1}{C_1(\bs)}\frac{\del C_0(\bs)}{\del a}\right]_{\bs=1}.
 \label{eq:sophisticated1}
\end{equation}
With \eqref{eq_A_1st_derivative} and \eqref{eq:sophisticated1} one is able to calculate the first moments \eqref{eq:A_moments_1st_tau} and \eqref{eq:A_moments_1st_jump}.

Similarly, after some extended algebra we obtain
\begin{multline}
 \left.\frac{\del^2\lambda_{\max}(\bs)}{\del a\del b}\right|_{\bs=1}=
 \left.\frac{1}{C_1}\frac{\del^2 C_0}{\del a\del b}\right|_{\bs=1}\\+\left[\frac{2C_2}{C_1^3}\frac{\del C_0}{\del a}\frac{\del C_0}{\del b}-\frac{1}{C_1^2}\left(\frac{\del C_0}{\del a}\frac{\del C_1}{\del b}+\frac{\del C_0}{\del b}\frac{\del C_1}{\del a}\right)\right]_{\bs=1}.\\
 \label{eq:sophisticated2}
\end{multline}
With this method, similar to \cite{koza99,bara15a,wach15}, one obtains for $i\neq j$ with $a=s_{ij}$ and $b=s_{kk}$ the coupled dispersion between a jump variable $\hat{n}_{ij}$ and the persistence time $\hat{\tau}_k$.
The procedure is as follows. One needs \eqref{eq:sophisticated1} and \eqref{eq:sophisticated2}
to calculate \eqref{eq_A_1st_derivative}, \eqref{eq_A_2nd_derivative},
which in turn with \eqref{eq:A_moments_1st_tau}-\eqref{eq:A_moments_2nd_jump}
can be used to calculate the covariances of persistence time and/or jump events.

To illustrate the procedure, let us use our specific two state model with modified generator \eqref{eq:A_mod_gen}
that leads to the characteristic polynomial
\begin{align}
 \chi_{\tilde{\mathcal{L}}(\bs)}(\lambda)&=w_{01}(x)w_{10}(x)(s_{00}s_{11}-s_{10}s_{01})\nonumber\\
 &+[w_{01}(x)s_{00}+w_{10}(x)s_{11}]\lambda+\lambda^2,
\end{align}
where we can identify $C_0(\bs)$ in the first line, $C_1$ in minus the brackets $C_1=-[\ldots]$, and $C_2=1$.
Specifically, for the first moment $\avg{\hat{\tau}_1}$, we set $a=s_{11}$ to get $C_1|_{\bs=1}=-[w_{01}(x)+w_{10}(x)]$
and $(\del_a C_0)_{\bs=1}=w_{01}(x)w_{10}(x)$. Using \eqref{eq:sophisticated1} implies $\del_a\lambda_\mathrm{max}(\bs)|_{\bs=1}=-w_{01}(x)w_{10}(x)/[w_{01}(x)+w_{10}(x)]$. Inserting the result into \eqref{eq_A_1st_derivative} and using \eqref{eq:A_moments_1st_tau},
which states that $\avg{\tau_1}\approx t\del_a\lambda_\mathrm{max}(\bs)|_{\bs=1}/[-w_{10}(x)]$,
yields the first moment
\begin{equation}
\avg{\tau_1}_x\approx\frac{w_{01}(x)t}{w_{01}(x)+w_{10}(x)},
 \label{eq:A_example1}
\end{equation}
which agrees with \eqref{eq:A_moment1a}.

For the variance $\avg{\hat{\tau}_1^2}_x-\avg{\hat{\tau}_1}_x^2$, we use \eqref{eq:sophisticated2} with $a=b=s_{11}$, $(\del_a^2C_0)_{\bs=1}=0$, $(\del_aC_0)_{\bs=1}=w_{01}(x)w_{10}(x)$,
and $C_1|_{\bs=0}=-[w_{01}(x)+w_{10}(x)]$, and $=(\del_a C_1)_{\bs=1}=-w_{10}(x)$,
to obtain
\begin{align}
 \frac{\del^2\lambda_\mathrm{max}}{\del a^2}&=
 \frac{2[w_{01}(x)w_{10}(x)]^2}{-[w_{01}(x)+w_{10}(x)]^3}+\frac{2w_{01}(x)w_{10}(x)^2}{[w_{01}(x)+w_{10}(x)]^2}\nonumber\\
 &=\frac{2w_{01}(x)w_{10}(x)^3}{[w_{01}(x)+w_{10}(x)]^3}.
\end{align}
With Eqs. \eqref{eq:A_moments_1st_tau} and \eqref{eq:A_moments_2nd_tau},
we obtain
\begin{equation}
 \avg{\hat{\tau}_1^2}-\avg{\hat{\tau}_1}^2\approx \frac{2w_{01}(x)w_{10}(x) t}{[w_{01}(x)+w_{10}(x)]^3},
\end{equation}
which agrees with \eqref{eq:A_Sigma_m|x}. Similarly, one can determine all remaining entries in the covariance matrix \eqref{eq:A_Sigma_m|x}.

\section{Optimal filtering for Gaussian process}
\label{sec:filter}
From the linear noise approximation, we have the Langevin equation
\begin{equation}
 \dot{\tilde{Y}}_t=-\omega_\y(\tilde{Y}_t-\avg{\tilde{Y}_\infty}_x)+\xi_t,
 \label{eq:A_langevin_lin}
\end{equation}
where $\omega_\y=w_{01}(0)+w_{10}(0)$, $\avg{\tilde{Y}_\infty}_x\equiv Np_0+Np_0(1-p_0)x$ with $p_0=w_{01}(0)/\omega_\y$, and finally, the white noise $\xi_t$ satisfies $\avg{\xi_t\xi_{t'}}=2D\delta(t-t')$ with $D\equiv N\omega_\y p_0(1-p_0)$.
The goal of filtering is to estimate the signal $x$
such that
\begin{equation}
 \tilde{\mathcal{E}}_t^2\equiv\avg{(x-\hat{x}_t)^2}
\end{equation}
becomes minimal, where $\hat{x}_t$ is a functional of the trajectory $\{\tilde{Y}_{t'}\}_{0\le t'\le t}$.
For a Gaussian process like \eqref{eq:A_langevin_lin}, one can verify that $\hat{x}_t$ must be a linear projection of $x$ on  $\{\tilde{Y}_{t'}\}_{0\le t'\le t}$,
i.e., $\avg{(x-\hat{x}_t)\tilde{Y}_{t'}}=0$ for $0\le t'\le t$ \cite{okse03}.

After rewriting \eqref{eq:A_langevin_lin} as an Ito differential equation
\begin{equation}
 \dd\tilde{Y}_t=-\omega_\y(\tilde{Y}_t-\avg{\tilde{Y}_\infty}_x)\dd t+\dd \xi_t,
 \label{eq:A_langevin_dlin}
\end{equation}
where $\dd \xi_t$ is the increment of a Wiener Process with $\dd \xi_t^2=2 N \omega_\y p_0(1-p_0)\dd t$,
one obtains, by
applying a Gram-Schmidt procedure for the projection $\hat{x}_t$, the differential equation \cite{okse03}
\begin{align}
 \dd \hat{x}_t&=\frac{\avg{x\dd\tilde{Y}_t^\perp}}{\avg{(\dd\tilde{Y}_t^\perp)^2}}\dd\tilde{Y}_t^\perp\nonumber\\
 &=\frac{\tilde{\mathcal{E}}_t^2}{2}\left[N\omega_\y p_0(1-p_0)(x-\hat{x}_t)\dd t+\dd\xi_t\right],
\end{align}
with $\hat{x}_0\equiv\avg{x}=0$. This in turn leads to the so-called Riccati equation
\cite{horo14a,okse03,hart16}
\begin{equation}
 \frac{\dd }{\dd t}\tilde{\mathcal{E}}_t^2=-\frac{N}{2}\omega_\y p_0(1-p_0)\tilde{\mathcal{E}}_t^4
\end{equation}
with $\tilde{\mathcal{E}}_0=\avg{x^2}=\mathcal{E}_\x^2$.
One can immediately see that the solution is given by
\begin{equation}
 \frac{1}{\tilde{\mathcal{E}}_t^2}=\frac{1}{\mathcal{E}_\x^2}+\frac{N\omega_\y p_0(1-p_0)}{2}t.
\end{equation}
Calculating the mutual information between the signal $x$ and the trajectory $\{\tilde{Y}_{t'}\}_{0\le t'\le t}$,
which according to \eqref{eq:A_Igauss} reads
\begin{multline}
\mathcal{I}^\mathrm{lin}_N\equiv
 I[x{:}\{\tilde{Y}_{t'}\}_{0\le t'\le t}]=I[x{:}\hat{x}_t]=
 \frac{1}{2}\ln\frac{\mathcal{E}_\x^2}{\tilde{\mathcal{E}}_t^2}\\
 =\frac{1}{2}\ln\left[1+\frac{N}{2}p_0(1-p_0)\mathcal{E}_\x^2\omega_\y t\right]\approx\tilde{\mathcal{I}}_N,\quad
\end{multline}
where we set $\mathcal{E}_{\x|\m}\to\tilde{\mathcal{E}}_t$ and identified the result finally with $\tilde{\mathcal{I}}_N$ given in \eqref{eq:Itilde_N}.
Note that all intermediate steps are exact due to the Gaussian nature of the coarse grained process $\tilde{Y}_t$.


\begin{thebibliography}{48}%
\makeatletter
\providecommand \@ifxundefined [1]{%
 \@ifx{#1\undefined}
}%
\providecommand \@ifnum [1]{%
 \ifnum #1\expandafter \@firstoftwo
 \else \expandafter \@secondoftwo
 \fi
}%
\providecommand \@ifx [1]{%
 \ifx #1\expandafter \@firstoftwo
 \else \expandafter \@secondoftwo
 \fi
}%
\providecommand \natexlab [1]{#1}%
\providecommand \enquote  [1]{``#1''}%
\providecommand \bibnamefont  [1]{#1}%
\providecommand \bibfnamefont [1]{#1}%
\providecommand \citenamefont [1]{#1}%
\providecommand \href@noop [0]{\@secondoftwo}%
\providecommand \href [0]{\begingroup \@sanitize@url \@href}%
\providecommand \@href[1]{\@@startlink{#1}\@@href}%
\providecommand \@@href[1]{\endgroup#1\@@endlink}%
\providecommand \@sanitize@url [0]{\catcode `\\12\catcode `\$12\catcode
  `\&12\catcode `\#12\catcode `\^12\catcode `\_12\catcode `\%12\relax}%
\providecommand \@@startlink[1]{}%
\providecommand \@@endlink[0]{}%
\providecommand \url  [0]{\begingroup\@sanitize@url \@url }%
\providecommand \@url [1]{\endgroup\@href {#1}{\urlprefix }}%
\providecommand \urlprefix  [0]{URL }%
\providecommand \Eprint [0]{\href }%
\providecommand \doibase [0]{http://dx.doi.org/}%
\providecommand \selectlanguage [0]{\@gobble}%
\providecommand \bibinfo  [0]{\@secondoftwo}%
\providecommand \bibfield  [0]{\@secondoftwo}%
\providecommand \translation [1]{[#1]}%
\providecommand \BibitemOpen [0]{}%
\providecommand \bibitemStop [0]{}%
\providecommand \bibitemNoStop [0]{.\EOS\space}%
\providecommand \EOS [0]{\spacefactor3000\relax}%
\providecommand \BibitemShut  [1]{\csname bibitem#1\endcsname}%
\let\auto@bib@innerbib\@empty
\bibitem [{\citenamefont {Berg}\ and\ \citenamefont {Purcell}(1977)}]{berg77}%
  \BibitemOpen
  \bibfield  {author} {\bibinfo {author} {\bibfnamefont {H.~C.}\ \bibnamefont
  {Berg}}\ and\ \bibinfo {author} {\bibfnamefont {E.~M.}\ \bibnamefont
  {Purcell}},\ }\href {\doibase 10.1016/S0006-3495(77)85544-6} {\bibfield
  {journal} {\bibinfo  {journal} {Biophys. J.}\ }\textbf {\bibinfo {volume}
  {20}},\ \bibinfo {pages} {193} (\bibinfo {year} {1977})}\BibitemShut
  {NoStop}%
\bibitem [{\citenamefont {Endres}\ and\ \citenamefont
  {Wingreen}(2009)}]{endr09}%
  \BibitemOpen
  \bibfield  {author} {\bibinfo {author} {\bibfnamefont {R.~G.}\ \bibnamefont
  {Endres}}\ and\ \bibinfo {author} {\bibfnamefont {N.~S.}\ \bibnamefont
  {Wingreen}},\ }\href {\doibase 10.1103/PhysRevLett.103.158101} {\bibfield
  {journal} {\bibinfo  {journal} {Phys. Rev. Lett.}\ }\textbf {\bibinfo
  {volume} {103}},\ \bibinfo {pages} {158101} (\bibinfo {year}
  {2009})}\BibitemShut {NoStop}%
\bibitem [{\citenamefont {Mora}\ and\ \citenamefont {Wingreen}(2010)}]{mora10}%
  \BibitemOpen
  \bibfield  {author} {\bibinfo {author} {\bibfnamefont {T.}~\bibnamefont
  {Mora}}\ and\ \bibinfo {author} {\bibfnamefont {N.~S.}\ \bibnamefont
  {Wingreen}},\ }\href {\doibase 10.1103/PhysRevLett.104.248101} {\bibfield
  {journal} {\bibinfo  {journal} {Phys. Rev. Lett.}\ }\textbf {\bibinfo
  {volume} {104}},\ \bibinfo {pages} {248101} (\bibinfo {year}
  {2010})}\BibitemShut {NoStop}%
\bibitem [{\citenamefont {Aquino}\ \emph {et~al.}(2014)\citenamefont {Aquino},
  \citenamefont {Tweedy}, \citenamefont {Heinrich},\ and\ \citenamefont
  {Endres}}]{aqui14}%
  \BibitemOpen
  \bibfield  {author} {\bibinfo {author} {\bibfnamefont {G.}~\bibnamefont
  {Aquino}}, \bibinfo {author} {\bibfnamefont {L.}~\bibnamefont {Tweedy}},
  \bibinfo {author} {\bibfnamefont {D.}~\bibnamefont {Heinrich}}, \ and\
  \bibinfo {author} {\bibfnamefont {R.~G.}\ \bibnamefont {Endres}},\ }\href
  {\doibase 10.1038/srep05688} {\bibfield  {journal} {\bibinfo  {journal} {Sci.
  Rep.}\ }\textbf {\bibinfo {volume} {4}},\ \bibinfo {pages} {5688} (\bibinfo
  {year} {2014})}\BibitemShut {NoStop}%
\bibitem [{\citenamefont {Hu}\ \emph {et~al.}(2010)\citenamefont {Hu},
  \citenamefont {Chen}, \citenamefont {Rappel},\ and\ \citenamefont
  {Levine}}]{hu10}%
  \BibitemOpen
  \bibfield  {author} {\bibinfo {author} {\bibfnamefont {B.}~\bibnamefont
  {Hu}}, \bibinfo {author} {\bibfnamefont {W.}~\bibnamefont {Chen}}, \bibinfo
  {author} {\bibfnamefont {W.-J.}\ \bibnamefont {Rappel}}, \ and\ \bibinfo
  {author} {\bibfnamefont {H.}~\bibnamefont {Levine}},\ }\href {\doibase
  10.1103/PhysRevLett.105.048104} {\bibfield  {journal} {\bibinfo  {journal}
  {Phys. Rev. Lett.}\ }\textbf {\bibinfo {volume} {105}},\ \bibinfo {pages}
  {048104} (\bibinfo {year} {2010})}\BibitemShut {NoStop}%
\bibitem [{\citenamefont {Mora}(2015)}]{mora15}%
  \BibitemOpen
  \bibfield  {author} {\bibinfo {author} {\bibfnamefont {T.}~\bibnamefont
  {Mora}},\ }\href {\doibase 10.1103/PhysRevLett.115.038102} {\bibfield
  {journal} {\bibinfo  {journal} {Phys. Rev. Lett.}\ }\textbf {\bibinfo
  {volume} {115}},\ \bibinfo {pages} {038102} (\bibinfo {year}
  {2015})}\BibitemShut {NoStop}%
\bibitem [{\citenamefont {Aquino}\ \emph {et~al.}(2016)\citenamefont {Aquino},
  \citenamefont {Wingreen},\ and\ \citenamefont {Endres}}]{aqui16}%
  \BibitemOpen
  \bibfield  {author} {\bibinfo {author} {\bibfnamefont {G.}~\bibnamefont
  {Aquino}}, \bibinfo {author} {\bibfnamefont {N.~S.}\ \bibnamefont
  {Wingreen}}, \ and\ \bibinfo {author} {\bibfnamefont {R.~G.}\ \bibnamefont
  {Endres}},\ }\href {\doibase 10.1007/s10955-015-1412-9} {\bibfield  {journal}
  {\bibinfo  {journal} {J. Stat. Phys.}\ }\textbf {\bibinfo {volume} {162}},\
  \bibinfo {pages} {1353} (\bibinfo {year} {2016})}\BibitemShut {NoStop}%
\bibitem [{\citenamefont {ten Wolde}\ \emph {et~al.}(2016)\citenamefont {ten
  Wolde}, \citenamefont {Becker}, \citenamefont {Ouldridge},\ and\
  \citenamefont {Mugler}}]{wold16}%
  \BibitemOpen
  \bibfield  {author} {\bibinfo {author} {\bibfnamefont {P.~R.}\ \bibnamefont
  {ten Wolde}}, \bibinfo {author} {\bibfnamefont {N.~B.}\ \bibnamefont
  {Becker}}, \bibinfo {author} {\bibfnamefont {T.~E.}\ \bibnamefont
  {Ouldridge}}, \ and\ \bibinfo {author} {\bibfnamefont {A.}~\bibnamefont
  {Mugler}},\ }\href {\doibase 10.1007/s10955-015-1440-5} {\bibfield  {journal}
  {\bibinfo  {journal} {J. Stat. Phys.}\ }\textbf {\bibinfo {volume} {162}},\
  \bibinfo {pages} {1395} (\bibinfo {year} {2016})}\BibitemShut {NoStop}%
\bibitem [{\citenamefont {Govern}\ and\ \citenamefont {ten
  Wolde}(2012)}]{gove12}%
  \BibitemOpen
  \bibfield  {author} {\bibinfo {author} {\bibfnamefont {C.~C.}\ \bibnamefont
  {Govern}}\ and\ \bibinfo {author} {\bibfnamefont {P.~R.}\ \bibnamefont {ten
  Wolde}},\ }\href {\doibase 10.1103/PhysRevLett.109.218103} {\bibfield
  {journal} {\bibinfo  {journal} {Phys. Rev. Lett.}\ }\textbf {\bibinfo
  {volume} {109}},\ \bibinfo {pages} {218103} (\bibinfo {year}
  {2012})}\BibitemShut {NoStop}%
\bibitem [{\citenamefont {Mehta}\ and\ \citenamefont {Schwab}(2012)}]{meht12}%
  \BibitemOpen
  \bibfield  {author} {\bibinfo {author} {\bibfnamefont {P.}~\bibnamefont
  {Mehta}}\ and\ \bibinfo {author} {\bibfnamefont {D.~J.}\ \bibnamefont
  {Schwab}},\ }\href {\doibase 10.1073/pnas.1207814109} {\bibfield  {journal}
  {\bibinfo  {journal} {Proc. Natl. Acad. Sci. USA}\ }\textbf {\bibinfo
  {volume} {109}},\ \bibinfo {pages} {17978} (\bibinfo {year}
  {2012})}\BibitemShut {NoStop}%
\bibitem [{\citenamefont {Tu}(2008)}]{tu08a}%
  \BibitemOpen
  \bibfield  {author} {\bibinfo {author} {\bibfnamefont {Y.}~\bibnamefont
  {Tu}},\ }\href {\doibase 10.1073/pnas.0804641105} {\bibfield  {journal}
  {\bibinfo  {journal} {Proc. Natl. Acad. Sci. USA}\ }\textbf {\bibinfo
  {volume} {105}},\ \bibinfo {pages} {11737} (\bibinfo {year}
  {2008})}\BibitemShut {NoStop}%
\bibitem [{\citenamefont {Lan}\ \emph {et~al.}(2012)\citenamefont {Lan},
  \citenamefont {Sartori}, \citenamefont {Neumann}, \citenamefont {Sourjik},\
  and\ \citenamefont {Tu}}]{lan12}%
  \BibitemOpen
  \bibfield  {author} {\bibinfo {author} {\bibfnamefont {G.}~\bibnamefont
  {Lan}}, \bibinfo {author} {\bibfnamefont {P.}~\bibnamefont {Sartori}},
  \bibinfo {author} {\bibfnamefont {S.}~\bibnamefont {Neumann}}, \bibinfo
  {author} {\bibfnamefont {V.}~\bibnamefont {Sourjik}}, \ and\ \bibinfo
  {author} {\bibfnamefont {Y.}~\bibnamefont {Tu}},\ }\href {\doibase
  10.1038/nphys2276} {\bibfield  {journal} {\bibinfo  {journal} {Nat. Phys.}\
  }\textbf {\bibinfo {volume} {8}},\ \bibinfo {pages} {422} (\bibinfo {year}
  {2012})}\BibitemShut {NoStop}%
\bibitem [{\citenamefont {Barato}\ \emph {et~al.}(2013)\citenamefont {Barato},
  \citenamefont {Hartich},\ and\ \citenamefont {Seifert}}]{bara13a}%
  \BibitemOpen
  \bibfield  {author} {\bibinfo {author} {\bibfnamefont {A.~C.}\ \bibnamefont
  {Barato}}, \bibinfo {author} {\bibfnamefont {D.}~\bibnamefont {Hartich}}, \
  and\ \bibinfo {author} {\bibfnamefont {U.}~\bibnamefont {Seifert}},\ }\href
  {\doibase 10.1103/PhysRevE.87.042104} {\bibfield  {journal} {\bibinfo
  {journal} {Phys. Rev. E}\ }\textbf {\bibinfo {volume} {87}},\ \bibinfo
  {pages} {042104} (\bibinfo {year} {2013})}\BibitemShut {NoStop}%
\bibitem [{\citenamefont {De~Palo}\ and\ \citenamefont
  {Endres}(2013)}]{palo13}%
  \BibitemOpen
  \bibfield  {author} {\bibinfo {author} {\bibfnamefont {G.}~\bibnamefont
  {De~Palo}}\ and\ \bibinfo {author} {\bibfnamefont {R.~G.}\ \bibnamefont
  {Endres}},\ }\href {\doibase 10.1371/journal.pcbi.1003300} {\bibfield
  {journal} {\bibinfo  {journal} {PLoS Comput. Biol.}\ }\textbf {\bibinfo
  {volume} {9}},\ \bibinfo {pages} {e1003300} (\bibinfo {year}
  {2013})}\BibitemShut {NoStop}%
\bibitem [{\citenamefont {Skoge}\ \emph {et~al.}(2013)\citenamefont {Skoge},
  \citenamefont {Naqvi}, \citenamefont {Meir},\ and\ \citenamefont
  {Wingreen}}]{skog13}%
  \BibitemOpen
  \bibfield  {author} {\bibinfo {author} {\bibfnamefont {M.}~\bibnamefont
  {Skoge}}, \bibinfo {author} {\bibfnamefont {S.}~\bibnamefont {Naqvi}},
  \bibinfo {author} {\bibfnamefont {Y.}~\bibnamefont {Meir}}, \ and\ \bibinfo
  {author} {\bibfnamefont {N.~S.}\ \bibnamefont {Wingreen}},\ }\href {\doibase
  10.1103/PhysRevLett.110.248102} {\bibfield  {journal} {\bibinfo  {journal}
  {Phys. Rev. Lett.}\ }\textbf {\bibinfo {volume} {110}},\ \bibinfo {pages}
  {248102} (\bibinfo {year} {2013})}\BibitemShut {NoStop}%
\bibitem [{\citenamefont {Lang}\ \emph {et~al.}(2014)\citenamefont {Lang},
  \citenamefont {Fisher}, \citenamefont {Mora},\ and\ \citenamefont
  {Mehta}}]{lang14}%
  \BibitemOpen
  \bibfield  {author} {\bibinfo {author} {\bibfnamefont {A.~H.}\ \bibnamefont
  {Lang}}, \bibinfo {author} {\bibfnamefont {C.~K.}\ \bibnamefont {Fisher}},
  \bibinfo {author} {\bibfnamefont {T.}~\bibnamefont {Mora}}, \ and\ \bibinfo
  {author} {\bibfnamefont {P.}~\bibnamefont {Mehta}},\ }\href {\doibase
  10.1103/PhysRevLett.113.148103} {\bibfield  {journal} {\bibinfo  {journal}
  {Phys. Rev. Lett.}\ }\textbf {\bibinfo {volume} {113}},\ \bibinfo {pages}
  {148103} (\bibinfo {year} {2014})}\BibitemShut {NoStop}%
\bibitem [{\citenamefont {Govern}\ and\ \citenamefont {ten
  Wolde}(2014{\natexlab{a}})}]{gove14}%
  \BibitemOpen
  \bibfield  {author} {\bibinfo {author} {\bibfnamefont {C.~C.}\ \bibnamefont
  {Govern}}\ and\ \bibinfo {author} {\bibfnamefont {P.~R.}\ \bibnamefont {ten
  Wolde}},\ }\href {\doibase 10.1073/pnas.1411524111} {\bibfield  {journal}
  {\bibinfo  {journal} {Proc. Natl. Acad. Sci. USA}\ }\textbf {\bibinfo
  {volume} {111}},\ \bibinfo {pages} {17486} (\bibinfo {year}
  {2014}{\natexlab{a}})}\BibitemShut {NoStop}%
\bibitem [{\citenamefont {Govern}\ and\ \citenamefont {ten
  Wolde}(2014{\natexlab{b}})}]{gove14a}%
  \BibitemOpen
  \bibfield  {author} {\bibinfo {author} {\bibfnamefont {C.~C.}\ \bibnamefont
  {Govern}}\ and\ \bibinfo {author} {\bibfnamefont {P.~R.}\ \bibnamefont {ten
  Wolde}},\ }\href {\doibase 10.1103/PhysRevLett.113.258102} {\bibfield
  {journal} {\bibinfo  {journal} {Phys. Rev. Lett.}\ }\textbf {\bibinfo
  {volume} {113}},\ \bibinfo {pages} {258102} (\bibinfo {year}
  {2014}{\natexlab{b}})}\BibitemShut {NoStop}%
\bibitem [{\citenamefont {Barato}\ \emph {et~al.}(2014)\citenamefont {Barato},
  \citenamefont {Hartich},\ and\ \citenamefont {Seifert}}]{bara14b}%
  \BibitemOpen
  \bibfield  {author} {\bibinfo {author} {\bibfnamefont {A.~C.}\ \bibnamefont
  {Barato}}, \bibinfo {author} {\bibfnamefont {D.}~\bibnamefont {Hartich}}, \
  and\ \bibinfo {author} {\bibfnamefont {U.}~\bibnamefont {Seifert}},\ }\href
  {\doibase 10.1088/1367-2630/16/10/103024} {\bibfield  {journal} {\bibinfo
  {journal} {New J. Phys.}\ }\textbf {\bibinfo {volume} {16}},\ \bibinfo
  {pages} {103024} (\bibinfo {year} {2014})}\BibitemShut {NoStop}%
\bibitem [{\citenamefont {Sartori}\ \emph {et~al.}(2014)\citenamefont
  {Sartori}, \citenamefont {Granger}, \citenamefont {Lee},\ and\ \citenamefont
  {Horowitz}}]{sart14}%
  \BibitemOpen
  \bibfield  {author} {\bibinfo {author} {\bibfnamefont {P.}~\bibnamefont
  {Sartori}}, \bibinfo {author} {\bibfnamefont {L.}~\bibnamefont {Granger}},
  \bibinfo {author} {\bibfnamefont {C.~F.}\ \bibnamefont {Lee}}, \ and\
  \bibinfo {author} {\bibfnamefont {J.~M.}\ \bibnamefont {Horowitz}},\ }\href
  {\doibase 10.1371/journal.pcbi.1003974} {\bibfield  {journal} {\bibinfo
  {journal} {PLoS Comput. Biol.}\ }\textbf {\bibinfo {volume} {10}},\ \bibinfo
  {pages} {e1003974} (\bibinfo {year} {2014})}\BibitemShut {NoStop}%
\bibitem [{\citenamefont {Hartich}\ \emph {et~al.}(2015)\citenamefont
  {Hartich}, \citenamefont {Barato},\ and\ \citenamefont {Seifert}}]{hart15}%
  \BibitemOpen
  \bibfield  {author} {\bibinfo {author} {\bibfnamefont {D.}~\bibnamefont
  {Hartich}}, \bibinfo {author} {\bibfnamefont {A.~C.}\ \bibnamefont {Barato}},
  \ and\ \bibinfo {author} {\bibfnamefont {U.}~\bibnamefont {Seifert}},\ }\href
  {\doibase 10.1088/1367-2630/17/5/055026} {\bibfield  {journal} {\bibinfo
  {journal} {New J. Phys.}\ }\textbf {\bibinfo {volume} {17}},\ \bibinfo
  {pages} {055026} (\bibinfo {year} {2015})}\BibitemShut {NoStop}%
\bibitem [{\citenamefont {{Bo}}\ \emph {et~al.}(2015)\citenamefont {{Bo}},
  \citenamefont {{Del Giudice}},\ and\ \citenamefont {{Celani}}}]{bo15}%
  \BibitemOpen
  \bibfield  {author} {\bibinfo {author} {\bibfnamefont {S.}~\bibnamefont
  {{Bo}}}, \bibinfo {author} {\bibfnamefont {M.}~\bibnamefont {{Del Giudice}}},
  \ and\ \bibinfo {author} {\bibfnamefont {A.}~\bibnamefont {{Celani}}},\
  }\href {\doibase 10.1088/1742-5468/2015/01/P01014} {\bibfield  {journal}
  {\bibinfo  {journal} {J. Stat. Mech.}\ ,\ \bibinfo {pages} {P01014}}
  (\bibinfo {year} {2015})}\BibitemShut {NoStop}%
\bibitem [{\citenamefont {Ito}\ and\ \citenamefont {Sagawa}(2015)}]{ito15}%
  \BibitemOpen
  \bibfield  {author} {\bibinfo {author} {\bibfnamefont {S.}~\bibnamefont
  {Ito}}\ and\ \bibinfo {author} {\bibfnamefont {T.}~\bibnamefont {Sagawa}},\
  }\href {\doibase 10.1038/ncomms8498} {\bibfield  {journal} {\bibinfo
  {journal} {Nat. Commun.}\ }\textbf {\bibinfo {volume} {6}},\ \bibinfo {pages}
  {7498} (\bibinfo {year} {2015})}\BibitemShut {NoStop}%
\bibitem [{\citenamefont {Barato}\ and\ \citenamefont
  {Seifert}(2015{\natexlab{a}})}]{bara15}%
  \BibitemOpen
  \bibfield  {author} {\bibinfo {author} {\bibfnamefont {A.~C.}\ \bibnamefont
  {Barato}}\ and\ \bibinfo {author} {\bibfnamefont {U.}~\bibnamefont
  {Seifert}},\ }\href {\doibase 10.1103/PhysRevLett.114.158101} {\bibfield
  {journal} {\bibinfo  {journal} {Phys. Rev. Lett.}\ }\textbf {\bibinfo
  {volume} {114}},\ \bibinfo {pages} {158101} (\bibinfo {year}
  {2015}{\natexlab{a}})}\BibitemShut {NoStop}%
\bibitem [{\citenamefont {Barato}\ and\ \citenamefont
  {Seifert}(2015{\natexlab{b}})}]{bara15a}%
  \BibitemOpen
  \bibfield  {author} {\bibinfo {author} {\bibfnamefont {A.~C.}\ \bibnamefont
  {Barato}}\ and\ \bibinfo {author} {\bibfnamefont {U.}~\bibnamefont
  {Seifert}},\ }\href {\doibase 10.1103/PhysRevE.92.032127} {\bibfield
  {journal} {\bibinfo  {journal} {Phys. Rev. E}\ }\textbf {\bibinfo {volume}
  {92}},\ \bibinfo {pages} {032127} (\bibinfo {year}
  {2015}{\natexlab{b}})}\BibitemShut {NoStop}%
\bibitem [{\citenamefont {Mehta}\ \emph {et~al.}(2016)\citenamefont {Mehta},
  \citenamefont {Lang},\ and\ \citenamefont {Schwab}}]{meht16}%
  \BibitemOpen
  \bibfield  {author} {\bibinfo {author} {\bibfnamefont {P.}~\bibnamefont
  {Mehta}}, \bibinfo {author} {\bibfnamefont {A.~H.}\ \bibnamefont {Lang}}, \
  and\ \bibinfo {author} {\bibfnamefont {D.~J.}\ \bibnamefont {Schwab}},\
  }\href {\doibase 10.1007/s10955-015-1431-6} {\bibfield  {journal} {\bibinfo
  {journal} {J. Stat. Phys.}\ }\textbf {\bibinfo {volume} {162}},\ \bibinfo
  {pages} {1153} (\bibinfo {year} {2016})}\BibitemShut {NoStop}%
\bibitem [{\citenamefont {Lan}\ and\ \citenamefont {Tu}(2016)}]{lan16}%
  \BibitemOpen
  \bibfield  {author} {\bibinfo {author} {\bibfnamefont {G.}~\bibnamefont
  {Lan}}\ and\ \bibinfo {author} {\bibfnamefont {Y.}~\bibnamefont {Tu}},\
  }\href {\doibase 10.1088/0034-4885/79/5/052601} {\bibfield  {journal}
  {\bibinfo  {journal} {Rep. Prog. Phys.}\ }\textbf {\bibinfo {volume} {79}},\
  \bibinfo {pages} {052601} (\bibinfo {year} {2016})}\BibitemShut {NoStop}%
\bibitem [{\citenamefont {van Kampen}(2007)}]{kamp07}%
  \BibitemOpen
  \bibfield  {author} {\bibinfo {author} {\bibfnamefont {N.~G.}\ \bibnamefont
  {van Kampen}},\ }\href {\doibase 10.1016/B978-044452965-7/50006-4} {\emph
  {\bibinfo {title} {Stochastic Processes in Physics and Chemistry}}},\
  \bibinfo {edition} {3rd}\ ed.,\ North-Holland Personal Library\ (\bibinfo
  {publisher} {Elsevier},\ \bibinfo {address} {Amsterdam},\ \bibinfo {year}
  {2007})\BibitemShut {NoStop}%
\bibitem [{\citenamefont {Elf}\ and\ \citenamefont {Ehrenberg}(2003)}]{elf03}%
  \BibitemOpen
  \bibfield  {author} {\bibinfo {author} {\bibfnamefont {J.}~\bibnamefont
  {Elf}}\ and\ \bibinfo {author} {\bibfnamefont {M.}~\bibnamefont
  {Ehrenberg}},\ }\href {\doibase 10.1101/gr.1196503} {\bibfield  {journal}
  {\bibinfo  {journal} {Genome Research}\ }\textbf {\bibinfo {volume} {13}},\
  \bibinfo {pages} {2475} (\bibinfo {year} {2003})}\BibitemShut {NoStop}%
\bibitem [{\citenamefont {Tkačik}\ and\ \citenamefont
  {Walczak}(2011)}]{tkac11}%
  \BibitemOpen
  \bibfield  {author} {\bibinfo {author} {\bibfnamefont {G.}~\bibnamefont
  {Tkačik}}\ and\ \bibinfo {author} {\bibfnamefont {A.~M.}\ \bibnamefont
  {Walczak}},\ }\href {\doibase 10.1088/0953-8984/23/15/153102} {\bibfield
  {journal} {\bibinfo  {journal} {J. Phys.: Condens. Matter}\ }\textbf
  {\bibinfo {volume} {23}},\ \bibinfo {pages} {153102} (\bibinfo {year}
  {2011})}\BibitemShut {NoStop}%
\bibitem [{\citenamefont {Bressloff}(2014)}]{bres14}%
  \BibitemOpen
  \bibfield  {author} {\bibinfo {author} {\bibfnamefont {P.~C.}\ \bibnamefont
  {Bressloff}},\ }\href {\doibase 10.1007/978-3-319-08488-6} {\emph {\bibinfo
  {title} {Stochastic Processes in Cell Biology}}}\ (\bibinfo  {publisher}
  {Springer International Publishing},\ \bibinfo {address} {Cham},\ \bibinfo
  {year} {2014})\BibitemShut {NoStop}%
\bibitem [{\citenamefont {Horowitz}(2015)}]{horo15a}%
  \BibitemOpen
  \bibfield  {author} {\bibinfo {author} {\bibfnamefont {J.~M.}\ \bibnamefont
  {Horowitz}},\ }\href {\doibase 10.1063/1.4927395} {\bibfield  {journal}
  {\bibinfo  {journal} {J. Chem. Phys.}\ }\textbf {\bibinfo {volume} {143}},\
  \bibinfo {eid} {044111} (\bibinfo {year} {2015})}\BibitemShut {NoStop}%
\bibitem [{\citenamefont {Koza}(1999)}]{koza99}%
  \BibitemOpen
  \bibfield  {author} {\bibinfo {author} {\bibfnamefont {Z.}~\bibnamefont
  {Koza}},\ }\href {\doibase 10.1088/0305-4470/32/44/303} {\bibfield  {journal}
  {\bibinfo  {journal} {J. Phys. A: Math. Gen.}\ }\textbf {\bibinfo {volume}
  {32}},\ \bibinfo {pages} {7637} (\bibinfo {year} {1999})}\BibitemShut
  {NoStop}%
\bibitem [{\citenamefont {Wachtel}\ \emph {et~al.}(2015)\citenamefont
  {Wachtel}, \citenamefont {Vollmer},\ and\ \citenamefont {Altaner}}]{wach15}%
  \BibitemOpen
  \bibfield  {author} {\bibinfo {author} {\bibfnamefont {A.}~\bibnamefont
  {Wachtel}}, \bibinfo {author} {\bibfnamefont {J.}~\bibnamefont {Vollmer}}, \
  and\ \bibinfo {author} {\bibfnamefont {B.}~\bibnamefont {Altaner}},\ }\href
  {\doibase 10.1103/PhysRevE.92.042132} {\bibfield  {journal} {\bibinfo
  {journal} {Phys. Rev. E}\ }\textbf {\bibinfo {volume} {92}},\ \bibinfo
  {pages} {042132} (\bibinfo {year} {2015})}\BibitemShut {NoStop}%
\bibitem [{\citenamefont {Seifert}(2012)}]{seif12}%
  \BibitemOpen
  \bibfield  {author} {\bibinfo {author} {\bibfnamefont {U.}~\bibnamefont
  {Seifert}},\ }\href {\doibase 10.1088/0034-4885/75/12/126001} {\bibfield
  {journal} {\bibinfo  {journal} {Rep. Prog. Phys.}\ }\textbf {\bibinfo
  {volume} {75}},\ \bibinfo {pages} {126001} (\bibinfo {year}
  {2012})}\BibitemShut {NoStop}%
\bibitem [{\citenamefont {Koski}\ \emph
  {et~al.}(2014{\natexlab{a}})\citenamefont {Koski}, \citenamefont {Maisi},
  \citenamefont {Pekola},\ and\ \citenamefont {Averin}}]{kosk14a}%
  \BibitemOpen
  \bibfield  {author} {\bibinfo {author} {\bibfnamefont {J.~V.}\ \bibnamefont
  {Koski}}, \bibinfo {author} {\bibfnamefont {V.~F.}\ \bibnamefont {Maisi}},
  \bibinfo {author} {\bibfnamefont {J.~P.}\ \bibnamefont {Pekola}}, \ and\
  \bibinfo {author} {\bibfnamefont {D.~V.}\ \bibnamefont {Averin}},\ }\href
  {\doibase 10.1073/pnas.1406966111} {\bibfield  {journal} {\bibinfo  {journal}
  {Proc. Natl. Acad. Sci. USA}\ }\textbf {\bibinfo {volume} {111}},\ \bibinfo
  {pages} {13786} (\bibinfo {year} {2014}{\natexlab{a}})}\BibitemShut {NoStop}%
\bibitem [{\citenamefont {Koski}\ \emph
  {et~al.}(2014{\natexlab{b}})\citenamefont {Koski}, \citenamefont {Maisi},
  \citenamefont {Sagawa},\ and\ \citenamefont {Pekola}}]{kosk14}%
  \BibitemOpen
  \bibfield  {author} {\bibinfo {author} {\bibfnamefont {J.~V.}\ \bibnamefont
  {Koski}}, \bibinfo {author} {\bibfnamefont {V.~F.}\ \bibnamefont {Maisi}},
  \bibinfo {author} {\bibfnamefont {T.}~\bibnamefont {Sagawa}}, \ and\ \bibinfo
  {author} {\bibfnamefont {J.~P.}\ \bibnamefont {Pekola}},\ }\href {\doibase
  10.1103/PhysRevLett.113.030601} {\bibfield  {journal} {\bibinfo  {journal}
  {Phys. Rev. Lett.}\ }\textbf {\bibinfo {volume} {113}},\ \bibinfo {pages}
  {030601} (\bibinfo {year} {2014}{\natexlab{b}})}\BibitemShut {NoStop}%
\bibitem [{\citenamefont {Koski}\ \emph {et~al.}(2015)\citenamefont {Koski},
  \citenamefont {Kutvonen}, \citenamefont {Khaymovich}, \citenamefont
  {Ala-Nissila},\ and\ \citenamefont {Pekola}}]{kosk15}%
  \BibitemOpen
  \bibfield  {author} {\bibinfo {author} {\bibfnamefont {J.~V.}\ \bibnamefont
  {Koski}}, \bibinfo {author} {\bibfnamefont {A.}~\bibnamefont {Kutvonen}},
  \bibinfo {author} {\bibfnamefont {I.~M.}\ \bibnamefont {Khaymovich}},
  \bibinfo {author} {\bibfnamefont {T.}~\bibnamefont {Ala-Nissila}}, \ and\
  \bibinfo {author} {\bibfnamefont {J.~P.}\ \bibnamefont {Pekola}},\ }\href
  {\doibase 10.1103/PhysRevLett.115.260602} {\bibfield  {journal} {\bibinfo
  {journal} {Phys. Rev. Lett.}\ }\textbf {\bibinfo {volume} {115}},\ \bibinfo
  {pages} {260602} (\bibinfo {year} {2015})}\BibitemShut {NoStop}%
\bibitem [{\citenamefont {Strasberg}\ \emph {et~al.}(2013)\citenamefont
  {Strasberg}, \citenamefont {Schaller}, \citenamefont {Brandes},\ and\
  \citenamefont {Esposito}}]{stra13}%
  \BibitemOpen
  \bibfield  {author} {\bibinfo {author} {\bibfnamefont {P.}~\bibnamefont
  {Strasberg}}, \bibinfo {author} {\bibfnamefont {G.}~\bibnamefont {Schaller}},
  \bibinfo {author} {\bibfnamefont {T.}~\bibnamefont {Brandes}}, \ and\
  \bibinfo {author} {\bibfnamefont {M.}~\bibnamefont {Esposito}},\ }\href
  {\doibase 10.1103/PhysRevLett.110.040601} {\bibfield  {journal} {\bibinfo
  {journal} {Phys. Rev. Lett.}\ }\textbf {\bibinfo {volume} {110}},\ \bibinfo
  {pages} {040601} (\bibinfo {year} {2013})}\BibitemShut {NoStop}%
\bibitem [{\citenamefont {Cover}\ and\ \citenamefont {Thomas}(2006)}]{cove06}%
  \BibitemOpen
  \bibfield  {author} {\bibinfo {author} {\bibfnamefont {T.~M.}\ \bibnamefont
  {Cover}}\ and\ \bibinfo {author} {\bibfnamefont {J.~A.}\ \bibnamefont
  {Thomas}},\ }\href {\doibase 10.1002/047174882X} {\emph {\bibinfo {title}
  {Elements of information theory}}},\ \bibinfo {edition} {2nd}\ ed.\ (\bibinfo
   {publisher} {Wiley-Interscience},\ \bibinfo {address} {Hoboken, NJ},\
  \bibinfo {year} {2006})\BibitemShut {NoStop}%
\bibitem [{\citenamefont {Parrondo}\ \emph {et~al.}(2015)\citenamefont
  {Parrondo}, \citenamefont {Horowitz},\ and\ \citenamefont {Sagawa}}]{parr15}%
  \BibitemOpen
  \bibfield  {author} {\bibinfo {author} {\bibfnamefont {J.~M.~R.}\
  \bibnamefont {Parrondo}}, \bibinfo {author} {\bibfnamefont {J.~M.}\
  \bibnamefont {Horowitz}}, \ and\ \bibinfo {author} {\bibfnamefont
  {T.}~\bibnamefont {Sagawa}},\ }\href {\doibase 10.1038/nphys3230} {\bibfield
  {journal} {\bibinfo  {journal} {Nat. Phys.}\ }\textbf {\bibinfo {volume}
  {11}},\ \bibinfo {pages} {131} (\bibinfo {year} {2015})}\BibitemShut
  {NoStop}%
\bibitem [{\citenamefont {Polettini}\ \emph {et~al.}(2015)\citenamefont
  {Polettini}, \citenamefont {Wachtel},\ and\ \citenamefont
  {Esposito}}]{pole15}%
  \BibitemOpen
  \bibfield  {author} {\bibinfo {author} {\bibfnamefont {M.}~\bibnamefont
  {Polettini}}, \bibinfo {author} {\bibfnamefont {A.}~\bibnamefont {Wachtel}},
  \ and\ \bibinfo {author} {\bibfnamefont {M.}~\bibnamefont {Esposito}},\
  }\href {\doibase 10.1063/1.4935064} {\bibfield  {journal} {\bibinfo
  {journal} {J. Chem. Phys.}\ }\textbf {\bibinfo {volume} {143}},\ \bibinfo
  {pages} {184103} (\bibinfo {year} {2015})}\BibitemShut {NoStop}%
\bibitem [{\citenamefont {Øksendal}(2003)}]{okse03}%
  \BibitemOpen
  \bibfield  {author} {\bibinfo {author} {\bibfnamefont {B.}~\bibnamefont
  {Øksendal}},\ }\href {\doibase 10.1007/978-3-642-14394-6} {\emph {\bibinfo
  {title} {Stochastic Differential Equations}}}\ (\bibinfo  {publisher}
  {Springer},\ \bibinfo {address} {Berlin Heidelberg},\ \bibinfo {year}
  {2003})\BibitemShut {NoStop}%
\bibitem [{\citenamefont {Horowitz}\ and\ \citenamefont
  {Sandberg}(2014)}]{horo14a}%
  \BibitemOpen
  \bibfield  {author} {\bibinfo {author} {\bibfnamefont {J.~M.}\ \bibnamefont
  {Horowitz}}\ and\ \bibinfo {author} {\bibfnamefont {H.}~\bibnamefont
  {Sandberg}},\ }\href {\doibase 10.1088/1367-2630/16/12/125007} {\bibfield
  {journal} {\bibinfo  {journal} {New J. Phys.}\ }\textbf {\bibinfo {volume}
  {16}},\ \bibinfo {pages} {125007} (\bibinfo {year} {2014})}\BibitemShut
  {NoStop}%
\bibitem [{\citenamefont {Hartich}\ \emph {et~al.}(2016)\citenamefont
  {Hartich}, \citenamefont {Barato},\ and\ \citenamefont {Seifert}}]{hart16}%
  \BibitemOpen
  \bibfield  {author} {\bibinfo {author} {\bibfnamefont {D.}~\bibnamefont
  {Hartich}}, \bibinfo {author} {\bibfnamefont {A.~C.}\ \bibnamefont {Barato}},
  \ and\ \bibinfo {author} {\bibfnamefont {U.}~\bibnamefont {Seifert}},\ }\href
  {\doibase 10.1103/PhysRevE.93.022116} {\bibfield  {journal} {\bibinfo
  {journal} {Phys. Rev. E}\ }\textbf {\bibinfo {volume} {93}},\ \bibinfo
  {pages} {022116} (\bibinfo {year} {2016})}\BibitemShut {NoStop}%
\bibitem [{\citenamefont {Pietzonka}\ \emph {et~al.}(2016)\citenamefont
  {Pietzonka}, \citenamefont {Barato},\ and\ \citenamefont {Seifert}}]{piet16}%
  \BibitemOpen
  \bibfield  {author} {\bibinfo {author} {\bibfnamefont {P.}~\bibnamefont
  {Pietzonka}}, \bibinfo {author} {\bibfnamefont {A.~C.}\ \bibnamefont
  {Barato}}, \ and\ \bibinfo {author} {\bibfnamefont {U.}~\bibnamefont
  {Seifert}},\ }\href {\doibase 10.1103/PhysRevE.93.052145} {\bibfield
  {journal} {\bibinfo  {journal} {Phys. Rev. E}\ }\textbf {\bibinfo {volume}
  {93}},\ \bibinfo {pages} {052145} (\bibinfo {year} {2016})}\BibitemShut
  {NoStop}%
\bibitem [{\citenamefont {Gingrich}\ \emph {et~al.}(2016)\citenamefont
  {Gingrich}, \citenamefont {Horowitz}, \citenamefont {Perunov},\ and\
  \citenamefont {England}}]{ging16}%
  \BibitemOpen
  \bibfield  {author} {\bibinfo {author} {\bibfnamefont {T.~R.}\ \bibnamefont
  {Gingrich}}, \bibinfo {author} {\bibfnamefont {J.~M.}\ \bibnamefont
  {Horowitz}}, \bibinfo {author} {\bibfnamefont {N.}~\bibnamefont {Perunov}}, \
  and\ \bibinfo {author} {\bibfnamefont {J.~L.}\ \bibnamefont {England}},\
  }\href {\doibase 10.1103/PhysRevLett.116.120601} {\bibfield  {journal}
  {\bibinfo  {journal} {Phys. Rev. Lett.}\ }\textbf {\bibinfo {volume} {116}},\
  \bibinfo {pages} {120601} (\bibinfo {year} {2016})}\BibitemShut {NoStop}%
\bibitem [{\citenamefont {Cottle}(1974)}]{cott74}%
  \BibitemOpen
  \bibfield  {author} {\bibinfo {author} {\bibfnamefont {R.~W.}\ \bibnamefont
  {Cottle}},\ }\href {\doibase 10.1016/0024-3795(74)90066-4} {\bibfield
  {journal} {\bibinfo  {journal} {Linear Algebra Appl.}\ }\textbf {\bibinfo
  {volume} {8}},\ \bibinfo {pages} {189} (\bibinfo {year} {1974})}\BibitemShut
  {NoStop}%
\end{thebibliography}
 \end{document}